\documentclass[twocolumn,showpacs,preprintnumbers,amsmath,amssymb,floatfix]{revtex4}
\usepackage{graphicx}% Include figure files
\usepackage{dcolumn}% Align table columns on decimal point
\usepackage{bm}% bold math

\begin{document}

\preprint{APS/123-QED}

\title{How informative are spatial CA3 representations established by the dentate gyrus?}% Force line breaks with \\

\author{Erika Cerasti}
 \email{cerasti@sissa.it}
\author{Alessandro Treves}%
 \altaffiliation[Also at ]{Kavli Institute for Systems Neuroscience and
Centre for the Biology of Memory, NTNU, Trondheim, Norway}%Lines break automatically or can be forced with \\
 \email{ale@sissa.it}
\affiliation{SISSA, Cognitive Neuroscience Sector, Trieste, Italy}%

\date{\today}% It is always \today, today,

\begin{abstract}
In the mammalian hippocampus, the dentate gyrus (DG) is characterized by sparse and powerful unidirectional projections to CA3 pyramidal cells, the so-called mossy fibers. Mossy fiber synapses appear to duplicate, in terms of the information they convey, what CA3 cells already receive from entorhinal cortex layer II cells, which project both to the dentate gyrus and to CA3. Computational models of episodic memory have hypothesized that the function of the mossy fibers is to enforce a new, well separated pattern of activity onto CA3 cells, to represent a new memory, prevailing over the interference produced by the traces of older memories already stored on CA3 recurrent collateral connections. Can this hypothesis apply also to spatial representations, as described by recent neurophysiological recordings in rats? To address this issue quantitatively, we estimate the amount of information DG can impart on a new CA3 pattern of spatial activity, using both mathematical analysis and computer simulations of a simplified model. We confirm that, also in the spatial case, the observed sparse connectivity and level of activity are most appropriate for driving memory storage -- and not to initiate retrieval. Surprisingly, the model also indicates that even when DG codes just for space, much of the information it passes on to CA3 acquires a non-spatial and episodic character, akin to that of a random number generator. It is suggested that further hippocampal processing is required to make full spatial use of DG inputs.
\end{abstract}

\maketitle

\section{Introduction}
The hippocampus presents the same organizaton across mammals, and distinct ones in reptiles and in birds. A most prominent and intriguing feature of the mammalian hippocampus is the dentate gyrus (DG). As reviewed in \cite{Treves08}, the dentate gyrus is positioned as a sort of intermediate station in the information flow between the entorhinal cortex and the CA3 region of the hippocampus proper. Since CA3 receives also direct, {\em perforant path} connections from entorhinal cortex, the DG inputs to CA3, called {\em mossy fibers}, appear to essentially duplicate the information that CA3 can already receive directly from the source. What may be the function of such a duplication?

Within the view that the {\em recurrent} CA3 network operates as an autoassociative memory \cite{McMorris}, \cite{Roll89}, it has been suggested that the mossy fibers (MF) inputs are those that drive the storage 
of new representations, whereas the perforant path (PP) inputs relay the cue that initiates the retrieval of a previously stored representation, through {\em attractor dynamics}, due largely to recurrent connections (RC). Such a proposal is supported by a mathematical model which allows a rough estimate of the amount of information, in bits, that different inputs may impart to a new CA3 representation \cite{Treves92}. That model, however, is formulated in the Marr \cite{Marr71} framework of {\em discrete} memory states, each of which is represented by a single activity configuration or firing pattern. 

Conversely, the prediction that MF inputs may be important for storage and not for retrieval has received tentative experimental support from experiments with spatial tasks, either the Morris water maze \cite{Lassalle} or a dry maze \cite{LeeKes}. Two-dimensional spatial representations, to be compatible with the attractor dynamics scenario, require a multiplicity of memory states, which approximate a 2D continuous manifold, isomorphic to the spatial environment to be represented. Moreover, there has to be of course a multiplicity of manifolds, to represent distinct environments with complete remapping from one to the other \cite{Leutgeb05}. Attractor dynamics then occurs along the dimensions locally orthogonal to each manifold, as in the simplified ``multi-chart" model \cite{SamsoMcN97}, \cite{Battaglia98}, whereas tangentially one expects marginal stability, allowing for small signals related to the movement of the animal, reflecting changing sensory cues as well as path integration, to displace a ``bump" of activity on the manifold, as appropriate \cite{SamsoMcN97}, \cite{Stringer}.

Although the notion of a really continuous attractor manifold appears as a limit case, which can only be approximated by a network of finite size \cite{Tsody95}, \cite{Hama}, \cite{Papp}, \cite{Roudi08}, even the limit case raises the issue of how a 2D attractor manifold can be established. In the rodent hippocampus, the above theoretical suggestion and partial experimental evidence point at a dominant role of the dentate gyrus, but it has remained unclear how the dentate gyrus, with its MF projections to CA3, can drive the establishment not just of a discrete pattern of activity, as envisaged by \cite{Treves92}, but of an entire spatial representation, in its full 2D glory. This paper reports the analysis of a simplified mathematical model aimed at addressing this issue in a quantitative, information theoretical fashion.

Such an analysis would have been difficult even only a few years ago, before the experimental discoveries that largely clarified, in the rodent, the nature of the spatial representations in the regions that feed into CA3. First, roughly half of the entorhinal PP inputs, those coming from layer II of the {\em medial} portion of entorhinal cortex, were found to be in the form of {\em grid cells}, i.e. units that are activated when the animal is in one of multiple regions, arranged on a regular triangular grid \cite{Haft05}. Second, the sparse activity earlier described in DG granule cells \cite{Jung94} was found to be concentrated on cells also with multiple fields, but irregularly arranged in the environment \cite{Leutgeb07}. These discoveries can now inform a simplified mathematical model, which would have earlier been based on ill-defined assumptions. Third, over the last decade {\em neurogenesis} in the adult dentate gyrus has been established as a quantitatively constrained but still significant phenomenon, stimulating novel ideas about its functional role \cite{Aimone08}. The first and third of these phenomena will be considered in extended versions of our model, to be analysed elsewhere; here, we focus on the role of the multiple DG place fields in establishing novel CA3 representations.

\subsection{A simplified mathematical model}

The complete model considers the firing rate of a CA3 pyramidal cell, $\eta_{i}$, to 
be determined by the firing rates $\{{\eta}\}$ of other cells in CA3, which influence it through RC connections; by the firing rates $\{{\beta}\}$ of DG granule cells, which feed into it through MF connections; by the firing rates
$\{{\varphi}\}$ of layer II pyramidal cells in entorhinal cortex (medial and lateral), which project to CA3 through PP axons; and by various feedforward and feedback inhibitory units. A most important simplification is that the fine temporal dynamics, e.g. on theta and gamma time scales, is neglected altogether, so that with ``firing rate" we mean an average over a time of order the theta period. Information coding over shorter time scales requires a more complex analysis, which is left to future refinements of the model. 

For the different systems of connections, we assume the existence of anatomical synapses between any two cells to be represented by fixed binary matrices $\{c^{PP}\}, \{c^{MF}\}, \{c^{RC}\}$ taking 0 or 1 values, whereas the efficacy of those synapses
to be described by matrices $\{J^{PP}\}, \{J^{MF}\}, \{J^{RC}\}$. The effect of inhibition and of the current threshold for activating a cell are summarized into a subtractive term, of which we denote with $\tilde{T}$ the mean value across CA3 cells, and with $\tilde{\delta_i}$ the deviation from the mean for a particular cell $i$. 

Assuming finally a simple threshold-linear activation function \cite{Treves90} for the relation between the activating current and the output firing rate, we write
\begin{widetext}
\begin{equation}
\eta_{i}(\vec{x})=g\left[\sum_{l}c^{PP}_{il}J^{PP}_{il}\varphi_{l}(\vec{x}) +\sum_{j}c^{MF}_{ij}J^{MF}_{ij}\beta_{j}(\vec{x}) + \sum_{k}c^{RC}_{ik}J^{RC}_{ik}\eta_{k}(\vec{x})+\tilde{\delta_i}-\tilde{T}\right]^{+}
\end{equation}
\end{widetext}
where $[\cdot]^{+}$ indicates taking the sum inside the brackets if positive in value, and zero if negative, and $g$ is a gain factor. The firing rates of the various populations are all assumed to depend on the position $\vec{x}$ of the animal, and the notation is chosen to minimize differences with our previous analyses of other components of the hippocampal system (e.g. \cite{Treves90}, \cite{Kropff08}).
 
\subsection{The storage of a new representation}

When the animal is exposed to a new environment, we make the drastic modelling
assumption that the new CA3 representation be driven solely by MF inputs, while PP and RC inputs provide interfering information, reflecting the storage of previous representations on those synaptic systems, i.e., noise. We reabsorb the
mean of such noise into the mean of the ``threshold+inhibition" term $\tilde{T}$ and similarly for the deviation from the mean, writing
\begin{equation}
\eta_{i}(\vec{x})=\left[\sum_{j}c^{MF}_{ij}J^{MF}_{ij}\beta_{j}(\vec{x})+\delta_i -T\right]^{+}
\label{firetaequ}
\end{equation}
where the gain has been set to $g=1$, without loss of generality, by an appropriate choice of the units in which to measure $\{c^{MF}\}, \{J^{MF}\}$ (pure numbers) and $\delta_i, T$ ($s^{-1}$).

As for the MF inputs, we consider a couple of simplified models that capture the essential finding by \cite{Leutgeb07}, of the irregularly arranged multiple fields, as well as the observed low activity level of DG granule cells \cite{Chawla}, while retaining the mathematical simplicity that favours an analytical treatment. We thus assume that only a randomly selected fraction $p_{DG}$ of the granule cells are active in a new environment, of size $A$, and that those units are active in a variable number $Q_j$ of locations, with $Q_j$ drawn from a distribution with mean $q$. In model A the distribution is taken to be Poisson (the data reported by Leutgeb et al \cite{Leutgeb07} are fit very well by a Poisson distribution with $q=1.7$, but their sampling is limited). In model B the distribution is taken to be exponential (this better describes the results of the simulations in \cite{Si09}, though that simple model may well be inappropriate). Therefore, in either model, the firing rate $\beta_j(\vec{x})$ of DG unit $j$ is a combination of $Q_j$ gaussian ``bumps", or fields, of equal effective size $(\sigma_f)^2$ and equal height $\beta_0$, centered at random points $\vec{x}_{jk}$ in the new environment
\begin{equation}
\beta_{j}\left(\vec{x}\right)=\sum_{k=0}^{Q_{j}}\; \beta_0\; e^{-\frac{\left(\vec{x}-\vec{x}_{jk}\right)^{2}}{2\sigma_{f}^{2}}}.
\end{equation}

The informative inputs driving the firing of a CA3 pyramidal cell, during storage of a new representation, result therefore from a combination of three distributions, in the model. The first, Poisson but close to normal, determines the MF connectivity, that is how it is that each CA3 unit receives only a few tens of connections out of $N_{DG}\simeq 10^6$ granule cells (in the rat), whereby $\left\{ c^{MF}_{ij} \right\}= 0,1$ with $ P\left( c^{MF}_{ij} =1 \right) =C_{MF}/N_{DG} \equiv c^{MF}$. The second, Poisson, determines which of the DG units presynaptic to a CA3 unit is active in the new environment, with $P(\mbox{unit}\; j \;\mbox{is active})=p_{DG}$. The third, either Poisson or exponential (and see model C below), determines how many fields an active DG unit has in the new environment. Note that in the rat $C_{MF} \simeq 46$ \cite{Amaral} whereas $p_{DG} \approx 0.02\div 0.05$, even when considering presumed newborn neurons \cite{Chawla}. As a result, the total number of active DG units presynaptic to a given CA3 unit, $p_{DG}C_{MF}\equiv \alpha$, is of order one, $\alpha\sim 1\div 2$, so that the second Poisson distribution effectively dominates over the first, and the number of active MF impinging on a CA3 unit can approximately be taken to be itself a Poisson variable with mean $\alpha$. As a qualification to such an approximation, one has to consider that different CA3 pyramidal cells, among the $N_{CA3}\simeq 3\times 10^5$ present in the rat (on each side), occasionally receive inputs from the {\em same} active DG granule cells, but rarely, as $N_{DG} \simeq 10^{6}$, hence the pool of active units $p_{DG}N_{DG}$ is only one order of magnitude smaller than the population of receiving units $N_{CA3}$.

In a further simplification, we consider the MF synaptic weights to be uniform in value, $J^{MF}_{ij}\equiv J$. This assumption, like those of equal height and width of the DG firing fields, is convenient for the analytical treatment but not necessary for the simulations. It will be relaxed later, in the computer simulations addressing the effect of MF synaptic plasticity. 

The new representation is therefore taken to be established by an informative signal coming from the dentate gyrus
\begin{equation}
\bar{\eta}_{i}(\vec{x})=J\sum_{j}c^{MF}_{ij}\beta_{j}(\vec{x})-T
\end{equation}
modulated, independently for each CA3 unit, by a noise term $\delta_i$, reflecting recurrent and perforant path inputs as well as other sources of variability, and which we take to be normally distributed with zero mean and standard deviation $\delta $.

The position $\vec{x}$ of the animal determines the firing $\{\beta\}$ of DG units, which in turn determine the probability 
distribution for the firing rate of any given CA3 pyramidal unit
$$
P\left(\eta_{i}|\vec{x}\right)=\delta(\eta_{i})\;\Phi\left(- \frac{\bar{\eta}_{i}(\vec{x})}{\delta}\right)
+\Theta(\eta_{i})\;\frac{e^{-\frac{\left(\eta_{i}-\bar{\eta}_{i}(\vec{x})\right)^{2}}{2\delta^{2}}}}{(\sqrt{2\pi}\delta)}\;
$$
where
$$
\Phi(\varrho(\vec{x}))\equiv \frac{1}{\sqrt{2\pi}}\int^{\varrho (\vec{x})}_{-\infty} e^{-t^{2}/2}dt
$$
is the integral of the gaussian noise up to given signal-to-noise ratio $$\varrho(\vec{x})\equiv\bar{\eta}(\vec{x})/\delta,$$ and $\Theta (\eta)$ is Heaviside's function vanishing for negative values of its argument. The first term, multiplying Dirac's $\delta\left(\eta_{i}\right)$, expresses the fact that negative activation values result in zero firing rates,
rather than negative rates.

Note that the resulting sparsity, i.e. how many of the CA3 units end up firing significantly at each position, which is a main 
factor affecting memory storage \cite{Treves90}, is determined by the threshold $T$, once the other parameters have been set. The approach taken here is to assume that the system requires the new representation to be sparse and regulates the threshold accordingly. We therefore set $a_{CA3}=0.1$, in broad agreement with experimental data \cite{Papp}, and adjust $T$ (as shown, for the mathematical analysis, in Sect.\ref{sparsity}).

The distribution of fields per DG unit is given in model A by the Poisson form
$$
P_A(Q)=\frac{q^{Q}}{Q!}e^{-q}
$$
in model B by the exponential form
$$
P_B(Q)=\frac{1}{1+q}\left(\frac{q}{1+q}\right)^{Q}
$$
and we also consider as a control case model C, where each DG unit has one and only one field
$$
P_C(Q)=\delta_{1Q}.
$$

\subsection{Assessing spatial information content}

In the model, spatial position $\vec{x}$ is represented by CA3 units, whose activity is informed about position by the activity of DG units, each determined independently of others by its place fields
$$
P\left(\{\beta(\vec{x})\}\right) = \prod_{j} P\left(\beta_{j}(\vec{x})\right)
$$
with
\begin{eqnarray}P\left(\beta_{j}(\vec{x})\right) &&= \;(1-p_{DG})\,\delta\left(\beta_{j}\right) + p_{DG}\times\nonumber \\
&&\sum_{Q_{j}=0}^{\infty}P_{A, B \; {\rm or} \; C}(Q_j)\; \delta\left(\beta_{j}-\sum^{Q_{j}}_{k=1}\psi\left(\vec{x}-\vec{x}_{jk}\right)\right)\nonumber
\end{eqnarray}
where each contributing field is a gaussian bump
$$
\psi\left(\vec{x}-\vec{x}_{jk}\right) \equiv
\left(\beta_0\; e^{-\frac{\left(\vec{x}-\vec{x}_{jk}\right)^{2}}{2\sigma_{f}^{2}}}
\right).
$$
The Mutual Information $I\left(\vec{x},\{\eta_{i}\}\right)$ quantifies the efficiency with which CA3 activity codes for position, on average, as
\begin{equation}
\langle I\left(\vec{x},\{\eta_{i}\}\right)\rangle\, =\,\langle H_{1}\left(\{\eta_{i}\}\right)\rangle - \langle \langle H_{2}\left(\{\eta_{i}\}|\vec{x}\right)\rangle_{\vec{x}}\rangle
\end{equation}
where the outer brackets $\langle \cdot\rangle$ indicate that the average is not just over the noise $\delta$, as usual in the estimation of mutual information, but also, in our case, over the {\em quenched}, i.e. constant but unknown values of the microscopic quantities
$c_{ij}$, the connectivity matrix, $Q_j$, the number of fields per active unit, and $\vec{x}_{jk}$, their centers.
For given values of the quenched variables, the total entropy $H_{1}$ and the (average) equivocation $H_{2}$ are defined as
\begin{eqnarray}
H_{1}\left(\{\eta_{i}\}\right)&=&-\int\prod_{i}d\eta_{i}\;P(\{\eta_{i}\})\log\left(P(\{\eta_{i}\})\right)\label{entr}\\
\langle H_{2}\left(\{\eta_{i}\}|\vec{x}\right)\rangle_{\vec{x}} &=&-\int (d\vec{x}/A)\prod_{i}d\eta_{i}\;P(\{\eta_{i}\}|\vec{x})\times \nonumber\\
&&\log\left(P(\{\eta_{i}\}|\vec{x})\right)\label{equiv}
\end{eqnarray}
where $A$ is the area of the given environment; the $\log$s are intended in base 2, to yield information values in bits.

The estimation of the mutual information can be approached analytically directly from these formulas, using the replica trick (see \cite{Mezard86}), as shown by \cite{Samengo} and \cite{DelPrete01}, and briefly described in Sect.\ref{replica}. As in those two studies, however, here too we are only able to complete the derivation in the limit of low signal-to-noise, or more precisely of limited variation, across space, of the signal-to-noise around its mean, that is 
$<(\varrho_i(\vec{x}) - < \varrho_i(\vec{x})>_{\vec{x}})^2 >_{\vec{x}}\to 0$. In this case we obtain, to first order in $N\equiv N_{CA3}$, an expression that can be shown to be equivalent to
\begin{widetext}
\begin{eqnarray}
\langle I\left(\vec{x},\{\eta_{i}\}\right)\rangle\, &=&\frac{N}{\ln2}\langle \int \frac{d\vec{x}}{A}
\left\{
\Phi(-\varrho_i(\vec{x}))\ln \Phi(-\varrho_i(\vec{x})) - \Phi(-\varrho_i(\vec{x}))\ln \int{\frac{d\vec{x}{\prime}}{A}\Phi(-\varrho_i(\vec{x}{\prime}))}\right\}\nonumber\\
&&\phantom{\frac{N}{\ln2}\langle}+ \int\frac{d\vec{x}}{A}\frac{d\vec{x}{\prime}}{A}\left\{ \frac{\Phi(\varrho_i(\vec{x}))}{2}\left[\varrho_{i}\left(\vec{x}\right)
-\varrho_{i}\left(\vec{x}{\prime}\right)\right]^2+
\left[\varrho_i(\vec{x})-\varrho_i(\vec{x}{\prime})\right]\sigma(\varrho_i(\vec{x}))\right\}\nonumber\\
&&\phantom{\frac{N}{\ln2}\langle}- \int\frac{d\vec{x}}{A}\frac{d\vec{x}{\prime}}{A}\frac{d\vec{x}{\prime\prime}}{A} \frac{\Phi(\varrho_i(\vec{x}))}{4}\left[\varrho_{i}\left(\vec{x}{\prime}\right)
-\varrho_{i}\left(\vec{x}{\prime\prime}\right)\right]^2 \rangle
\label{info-space}
\end{eqnarray}
where we use the notation $\sigma(\varrho)=(1/\sqrt{2\pi})\exp{-\varrho^2/2}$ (cp. \cite{DelPrete01}, Eqs.17, 45).

Being limited to the first order in $N$, the expression above can be obtained in a straightforward manner by directly expanding the logarithms, in the large noise limit $\delta \to\infty$, in the simpler formula quantifying the information conveyed by a single CA3 unit
\begin{eqnarray}
\langle I\left(\vec{x},\{\eta_{i}\}\right)\rangle\, &=&\frac{1}{\ln2}\langle \int \frac{d\vec{x}}{A}
\left\{
\Phi(-\varrho_i(\vec{x}))\ln \Phi(-\varrho_i(\vec{x})) - \Phi(-\varrho_i(\vec{x}))\ln \int{\frac{d\vec{x}{\prime}}{A}\Phi(-\varrho_i(\vec{x}{\prime}))}\right\}\nonumber\\
&&\phantom{\frac{N}{\ln2}\langle}-\int \frac{d\vec{x}}{A}\int   \frac{d\eta}{\sqrt{2\pi}\delta}e^{-\frac{\left(\eta-\bar{\eta}(x)\right)^2}{2\delta^2}} \ln\left[ \int \frac{d\vec{y}}{A}\; e^{\frac{\bar{\eta}^2({x})-\bar{\eta}^2({y})-2\eta\left(\bar{\eta}(x)-\bar{\eta}(y)\right)}{2\delta^2}} \right]\rangle\nonumber\\
\label{info-space1cell}
\end{eqnarray}
\end{widetext}

This single-unit formula cannot quantify the higher-order contributions in $N$, which decrease the information conveyed by a population in which some of the units inevitably convey some of the same information. The replica derivation, instead, in principle would allow one to take into proper account such correlated selectivity, which ultimately results in the information conveyed by large CA3 populations not scaling up linearly with $N$, and saturating instead once enough CA3 units have been sampled, as shown in related models by \cite{Samengo}, \cite{DelPrete01}. In our case however the calculation of e.g. the second order terms in $N$ is further complicated by the fact that different CA3 units receive inputs coming from partially overlapping subsets of DG units. This may cause saturation at a lower level, once all DG units have been effectively sampled. The interested reader can follow the derivation sketched in Sect.\ref{replica}.

Having to take, in any case, the large noise limit implies that the resulting formula is not really applicable to neuronally plausible values of the parameters, but only to the uninteresting case in which DG units impart very little information onto CA3 units. Therefore we use only the single-unit formula, and resort to computer simulations to assess the effects of correlated DG inputs.
Sects. \ref{decomposition} and \ref{sparsity} indicate how to obtain numerical results by evaluating the expression in Eq.~\ref{info-space1cell}.

Computer simulations can be used to estimate the information present in samples of CA3 units of arbitrary size, and at arbitrary levels of noise, but at the price of an indirect {\em decoding} procedure. A decoding step is required because the dimensionality of the space spanned by the CA3 activity $\{\eta_{i}\}$ is too high. The decoding method we use, described in Sect.\ref{decoding}, leads to two different types of information estimates, based on either the full or reduced localization matrix. The difference between the two is illustrated under Results and further discussed at the end of the paper.

\section{Results}

The essential mechanism described by the model is very simple, as illustrated in Fig.1. CA3 units which happen to receive a few DG overlapping fields combine them in a resulting field of their own, that can survive thresholding. The devil is in the quantitative details: how often this occurs, how large are the output fields, how distinct above the noise, all factors that determine the information contained in the spatial representation. Note that the same CA3 unit can express multiple fields. 

It is convenient to discuss such quantitative details with reference to a standard set of parameters. Our model of reference is a network of DG units with fields representated by Gaussian-like functions of space. The number of fields per each DG units is given by a Poisson distribution with mean value $q$, and parameters as specified in Table \ref{table1}.

\begin{table*}
\begin{tabular}{|l|c|p{5cm}|}
\hline
parameter & symbol & standard value \\
\hline
probability a DG unit is active in one environment & $p_{DG}$ & 0.033 \\
\hline
mean number of DG inputs to a CA3 unit & $C_{MF}$ & 50 \\
\hline
mean number of fields per active DG unit & $q$ & 1.7 \\
\hline
mean number of fields activating a CA3 unit & $\mu$ & $C_{MF}p_{DG}q=2.833$ \\
\hline
strength of MF inputs & $J$ & 1, otherwise $2.833/\mu$ \\
\hline
noise affecting CA3 activity & $\delta$ & 1 (in units in which $\beta_0=2.02$)  \\
\hline
sparsity of CA3 activity & $a_{CA3}$ & 0.1  \\
\hline
\end{tabular}
\caption{Parameters used in the standard version of the model.\label{table1}}
\end{table*}

\begin{figure}
\includegraphics[width=0.8\linewidth]{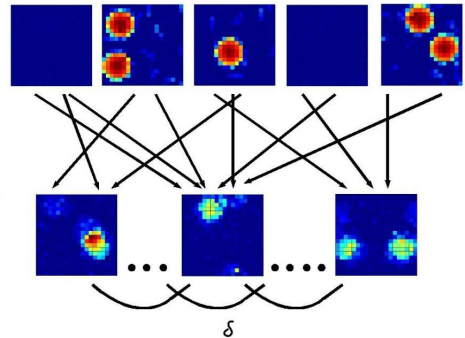}
\caption{{\bf Network scheme:} The DG-CA3 system indicating examples of the fields attributed to DG units and of those resulting in CA3 units, the connectivity between the two populations, and the noise $\delta$ that replaces, in the model, also the effect of recurrent connections in CA3.\label{fig:1}}
\end{figure}

In general, the stronger the mean DG input, the more it dominates over the noise, and also the higher the threshold has to be set in CA3 to make the pattern of activity as sparse as required. To control for the trivial advantage of a higher signal-to-noise, we perform comparisons in which it is kept fixed, by adjusting e.g. the MF synaptic strength $J$.

\subsection{Multiple input cells vs. multiple fields per cell}
The first parameter we considered is $q$, in light of the recent finding that DG units active in a restricted environment appear to 
have more often multiple fields than CA3 units, and much more often than expected, given their weak probability of being active 
\cite{Leutgeb07}. We wondered whether receiving multiple fields from the same input units would be advantageous for CA3, and if so whether there is an optimal $q$ value. We therefore estimated the mutual information when $q$ varies and $\mu$, the total mean number of DG fields that each CA3 cell receives as input, is kept fixed, by varying $C_{MF}$ correspondigly. Initially, we did indeed find
an optimal $q$ value, that appeared to maximize the information available in CA3, and the optimal value was consistent with the recordings of \cite{Leutgeb07}. After discovering a mistake in our initial analyses, however, we have realized that varying $q$ in this way makes very little difference. Fig.2 reports the results of computer simulations, that illustrate also the dependence of the mutual information on $N_{CA3}$, the number of cells sampled. The dependence is sub-linear, but rather smooth, with significant fluctuations from sample-to-sample which are largely averaged out in the graph. The different lines correspond to different distributions of the input DG fields among active DG cells projecting to CA3, that is different combinations of values for $q$ and $C_{MF}=\mu/(qp_{DG})$, with $\mu$ kept constant; these different distributions do not affect much the information in the representation. 

\begin{figure}
\includegraphics[width=\linewidth]{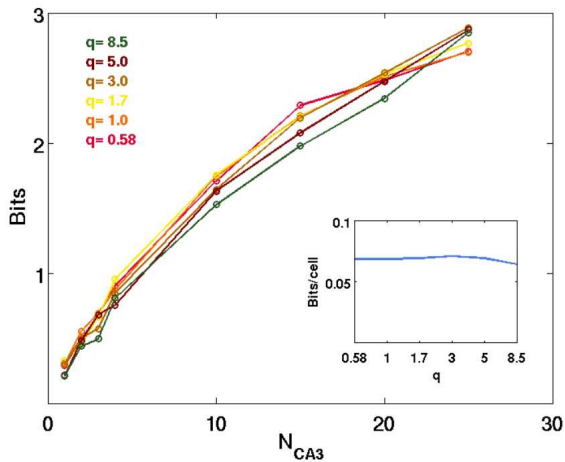}
\caption{{\bf The exact multiplicity of fields in DG units is irrelevant.} Information about position plotted versus the number of CA3 units from which it is decoded, with the mean number of fields in the input to each CA3 unit constant at the value $\mu=2.833$. Different lines correspond to a different mean number of fields per DG input units, balanced by different mean number of input units per CA3 unit. Inset: analytical estimate of the information per CA3 unit, from numerically integrating Eq.~\ref{info-space1cell}.}
\label{fig:2}
\end{figure}

The analytical estimate of the information per CA3 unit confirms that there is no dependence on $q$ (Fig.~\ref{fig:2}, inset). This is not a trivial result, as it would be if only the parameter $\mu$ entered the analytical expression. Instead, Sect.~\ref{decomposition} shows that the parameters $C_m$ of the $m$-field decomposition depend separately on $q$ and $\alpha\equiv p_{DG}C_{MF}$, so the fact that the two separate dependencies almost cancel out in a single dependence on their product, $\mu$, is remarkable. Moreover, such analytical estimate of the information conveyed by one unit does not match the first datapoints, for $N_{CA3}=1$, extracted from the computer simulation; it is not higher, as might have been expected considering that the simulation requires an additional information loosing decoding step, but lower, by over a factor of 2. The finding that the analytical estimate differs from, and is in fact much lower than, the slope parameter extracted from the simulations, after the decoding step, is further discussed below. What the simulations and the analytical estimate have in common, beyond their incongruence in absolute values, is the absence of separate dependencies on $q$ and $\alpha$, as shown in Fig.~\ref{fig:2}. 

\subsection{More MF connections, but weaker}
\begin{figure*}
\begin{tabular}{lc}
(a)\begin{minipage}{3.2in}
\includegraphics[width=\linewidth]{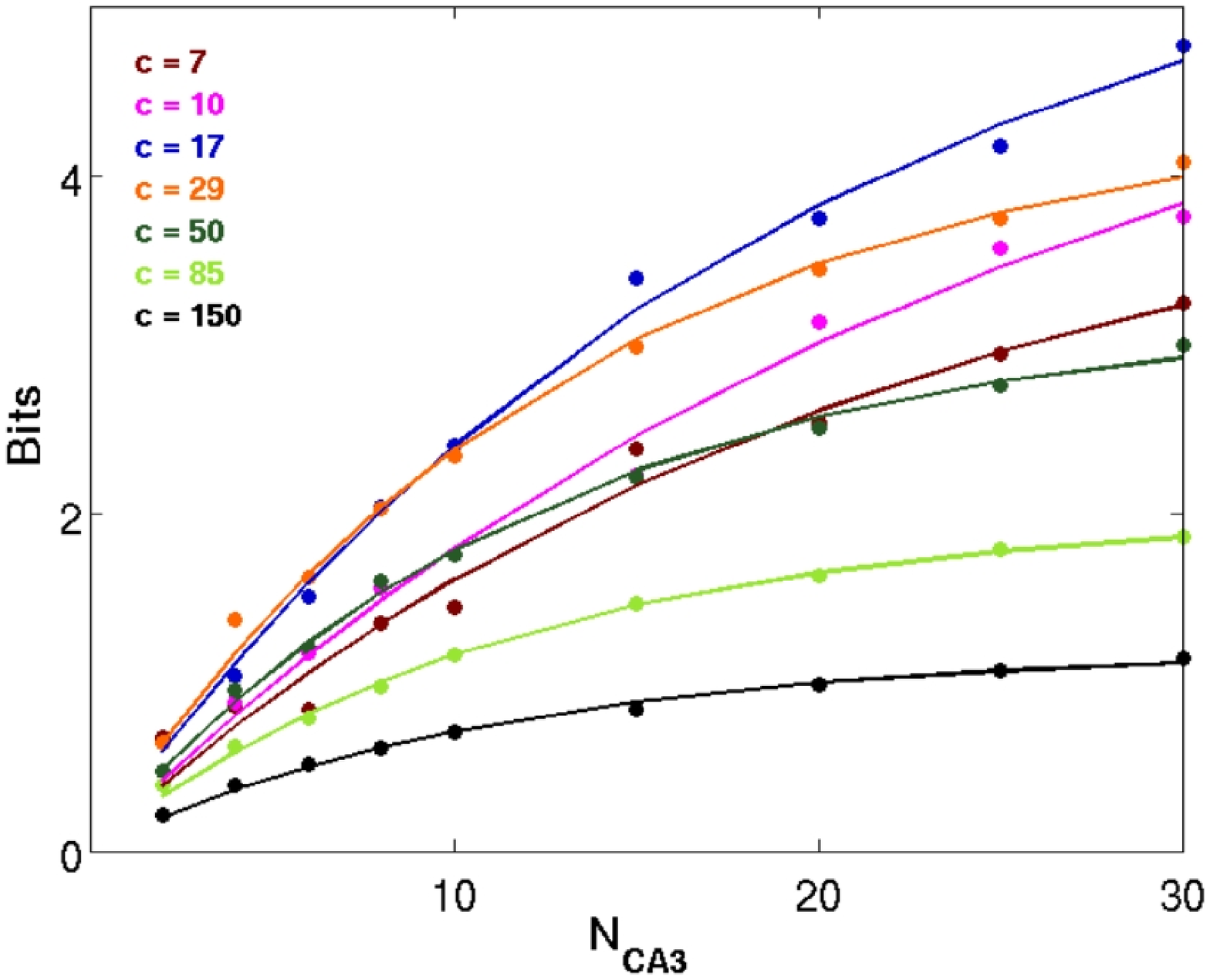}
\end{minipage}&
(b)\begin{minipage}{3.2in}
\includegraphics[width=\linewidth]{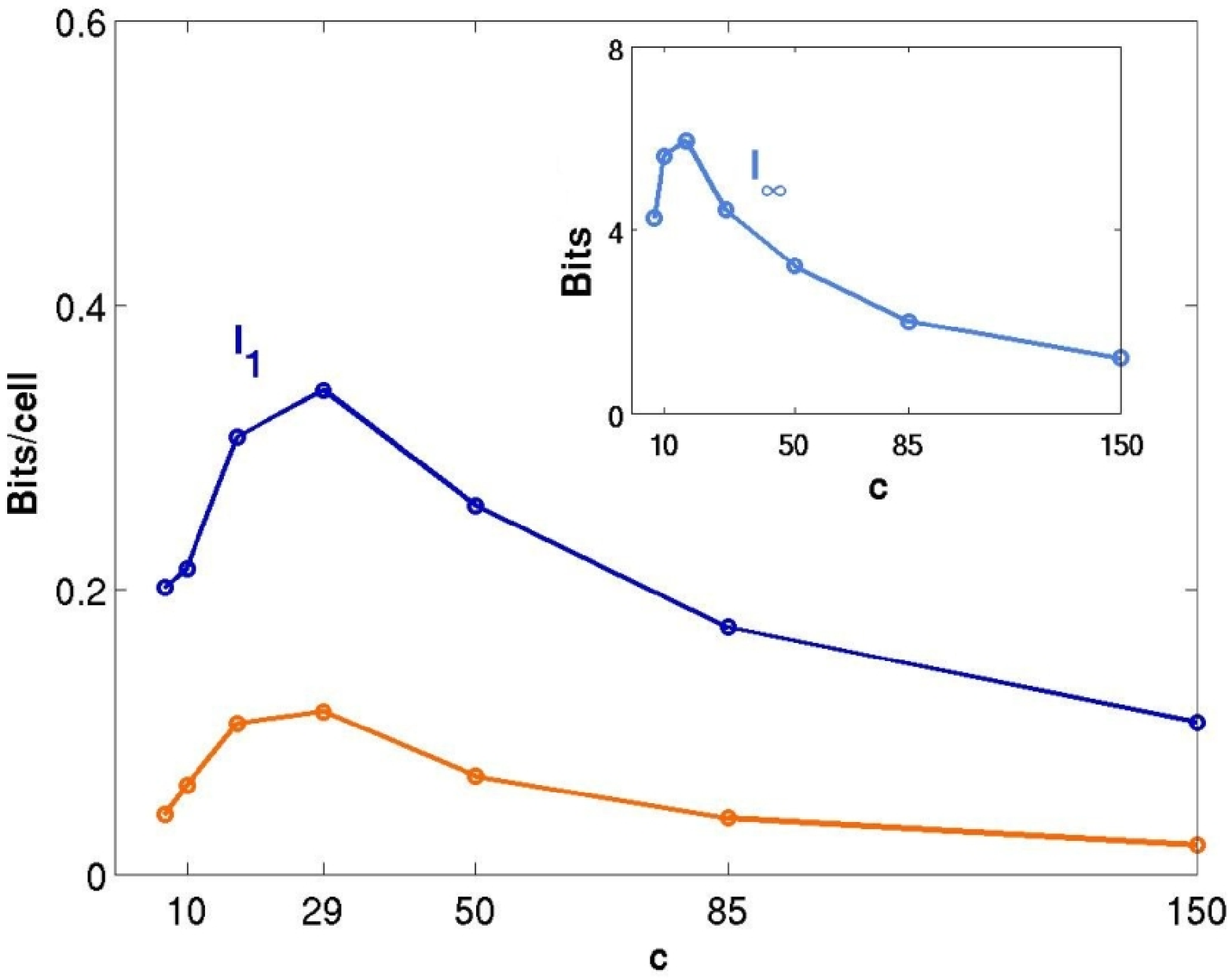}
\end{minipage}\\
\end{tabular}
\caption{{\bf A sparse MF connectivity is optimal, but not too sparse.} (a) Information plotted versus the number of CA3 cells, with different colors for different values of $C_{MF}$. Dots represent information values obtained from simulations, while curves are exponentially saturating fits to the data points, as described in Methods. (b) Plot of the two parameters of the fit curves. Main figure: slope parameter describing the slope of the linear part of the curve (for low $N_{CA3}$), constrasted with the analytical estimate of the term proportional to $N_{CA3}$ (Eq.\ref{info-space1cell}); inset: total information parameter, describing the saturation level reached by the curve.}\label{fig:3}
\end{figure*}

Motivated by the striking sparsity of MF connections, compared to the thousands of RC and PP synaptic connections impinging on CA3 cells in the rat, we have then tested the effect of changing $C_{MF}$ without changing $q$. In order to vary the mean number of DG units that project to a single CA3 unit, while keeping constant the total mean input strength, assumed to be an independent biophysically constrained parameter, we varied inversely to $C_{MF}$ the synaptic strength parameter $J$.
As shown in Fig.~\ref{fig:3}, the information presents a maximum at some intermediate value $C_{MF}\simeq 20-30$, which is observed both in simulations and in the analytical estimate, despite the fact that again they differ by more than a factor of two. 

Again we find that the analytical estimate differs from, and is in fact much lower than, the slope parameter extracted from the simulations, after the decoding ste. Both measures, however, show that the standard model is not indifferent to how sparse are the MF connections. If they are very sparse, most CA3 units receive no inputs from active DG units, and the competition induced by the sparsity constraint tends to be won, at any point in space, by those few CA3 units that are receiving input from just one active DG unit.  The resulting mapping is effectively one-to-one, unit-to-unit, and this is not optimal information-wise, because too few CA3 units are active -- many of them in fact have multiple fields (Fig.\ref{fig:3}b), reflecting the multiple fields of their ``parent" units in DG. As $C_{MF}$ increases (with a corresponding decrease in MF synaptic weight), the units that win the competition tend to be those that summate inputs from two or more concurrently active DG units. The mapping ceases to be one-to-one, and this increases the amount of information, up to a point. When $C_{MF}$ is large enough that CA3 units begin to sample more effectively DG activity, those that win the competition tend to be the ``happy few" that happen to summate several active DG inputs, and this tends to occur at only one place in the environment. As a result, an ever smaller fraction of CA3 units have place fields, and those tend to have just one, often very irregular, as shown in Fig.\ref{fig:3}b. From that point on, the information in the representation decreases monotonically. The optimal MF connectivity is then in the range which maximizes the fraction of CA3 units that have a field in the newly learned environment, at a value, roughly one third, broadly consistent with experimental data (see e.g. \cite{Leutgeb04}).

\begin{figure*}
\begin{tabular}{lc}
(a) \begin{minipage}{3.2in}
\includegraphics[width=\linewidth]{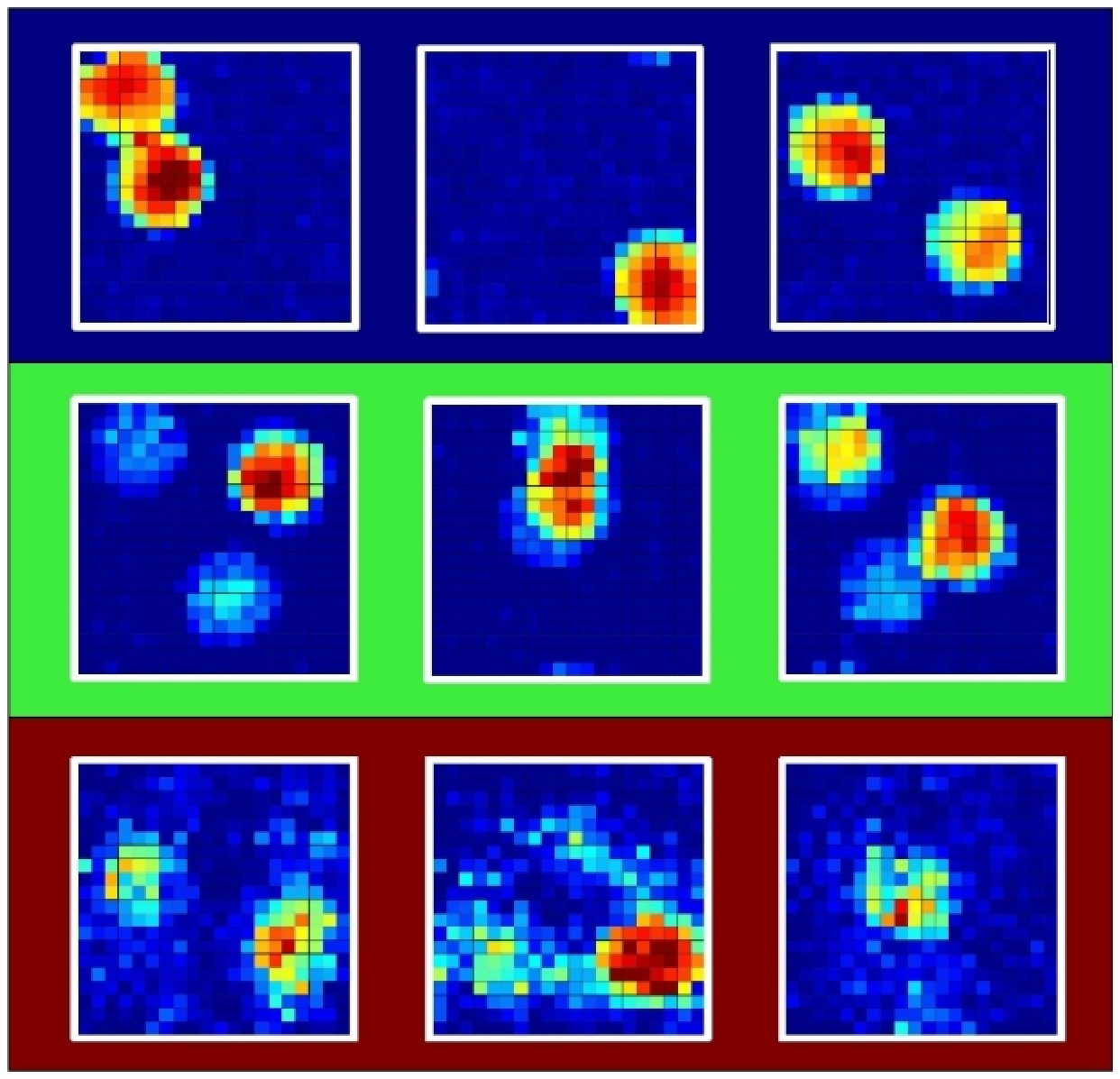}
\end{minipage}&
(b)\begin{minipage}{3.2in}
\includegraphics[width=\linewidth]{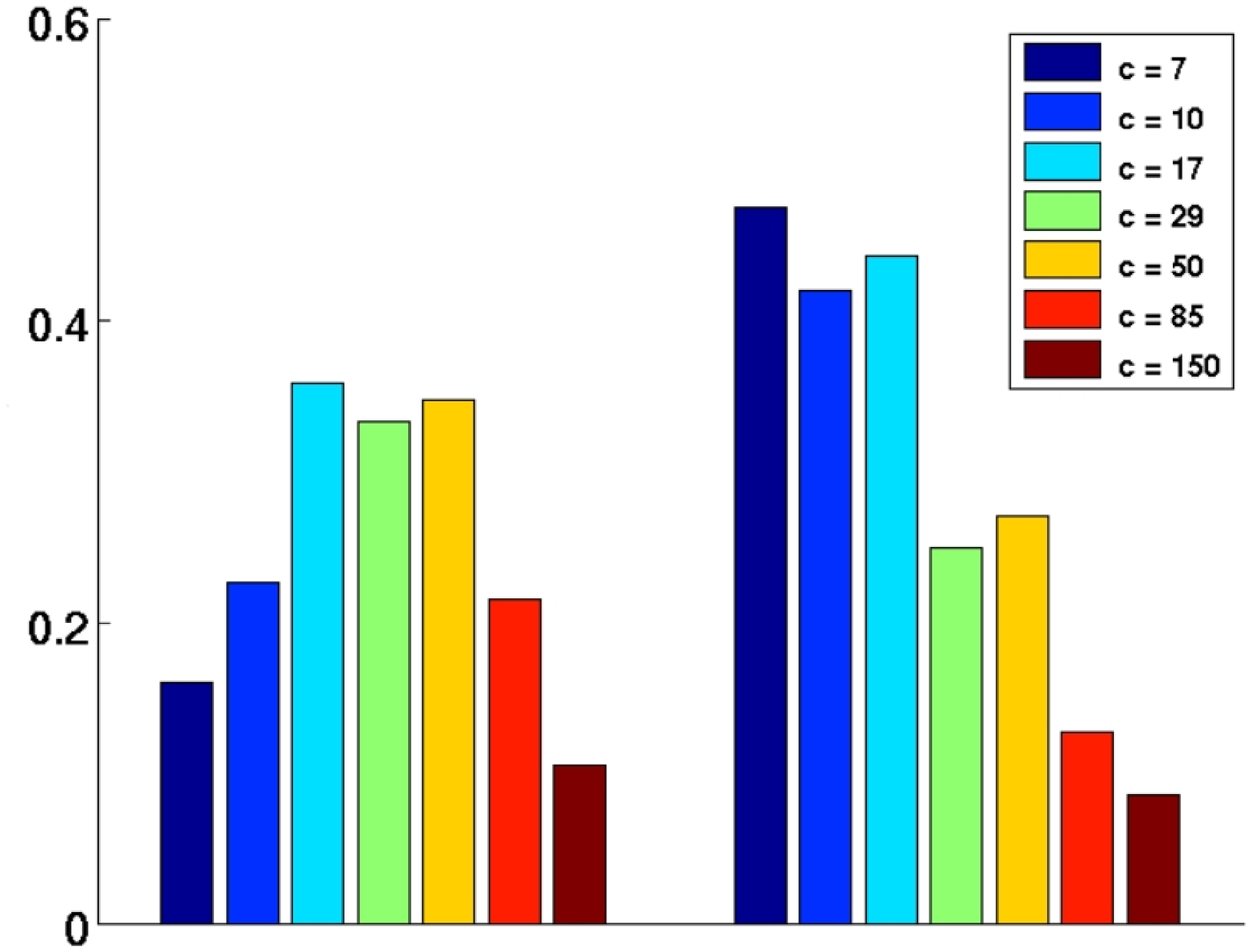}
\end{minipage}
\end{tabular}
\caption{ {\bf Information vs. connectivity:} (a) Examples of CA3 firing rate maps for $C_{MF}=7$ (top row); $C_{MF}=29$ (middle) and $C_{MF}=150$ (bottom); (b) Histogram that shows the fraction of CA3 units active somewhere in the environment, left, and the fraction of these with more than one field, right, for different $C_{MF}$ values.}\label{fig:4}
\end{figure*}

It is important to emphasize that what we are reporting is a quantitative effect: the underlying mechanism is always the same, the random summation of inputs from active DG units. DG in the model effectively operates as a sort of random number generator, whatever the values of the various parameters. How informative are the CA3 representations established by that random number generator, however, depends on the values of the parameters.

\subsection{Other DG field distribution models}

\begin{figure}
\centering
\includegraphics[width=\linewidth]{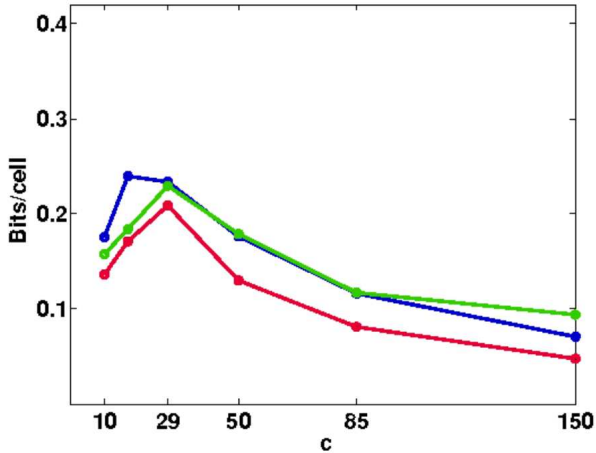}
\caption{{\bf Information vs. connectivity:} Information plotted versus different values of connectivity between DG and CA3. Solid lines are all from simulations (localization information from samples of 10 units, divided by 10), as follows: for the blue line, the distribution defining the number of fields in DG cells is Poisson (model A); for the green line, it is exponential (model B); and for the red line, each DG active unit has one field only (model C).}\label{fig:5}
\end{figure}

We repeated the simulations using other models for the DG fields distribution, the exponential (model B) and the single field one (model C), and the results are similar to those obtained for model A: the information has a maximum when varying $C_{MF}$ on its own, and is instead roughly constant if the parameter $\mu$ is held constant (by varying $q$ inversely to $C_{MF}$). Fig.~\ref{fig:5} reports the comparison, as $C_{MF}$ varies, between models A and B, with $q=1.7$, and model C, where $q\equiv 1$, so that in this latter case the inputs are 1/1.7 times weaker (we did not compensate by multiplying $J$ by 1.7). Information measures are obtained by decoding several samples of 10 units, averaging and dividing by 10, and not by extracting the fit parameters. As one can see, the lower mean input for model C leads to lower information values, but the trend with $C_{MF}$ is the same in all three models. This further indicates that the multiplicity of fields in DG units, as well as its exact distribution, is of no major consequence, if comparisons are made keeping constant the mean number of fields in the input to a CA3 unit. 

\subsection{Sparsity of DG activity}
We study also how the level of DG activity affects the information flow. We choose diffferent values for the probability $p_{DG}$ that a single DG unit fires in the given environment, and again we adjust the synaptic weight $J$ to keep the mean DG input per CA3 cell constant across the comparisons.

\begin{figure*}
\begin{tabular}{lc}
(a) \begin{minipage}{3.2in}
\includegraphics[width=\linewidth]{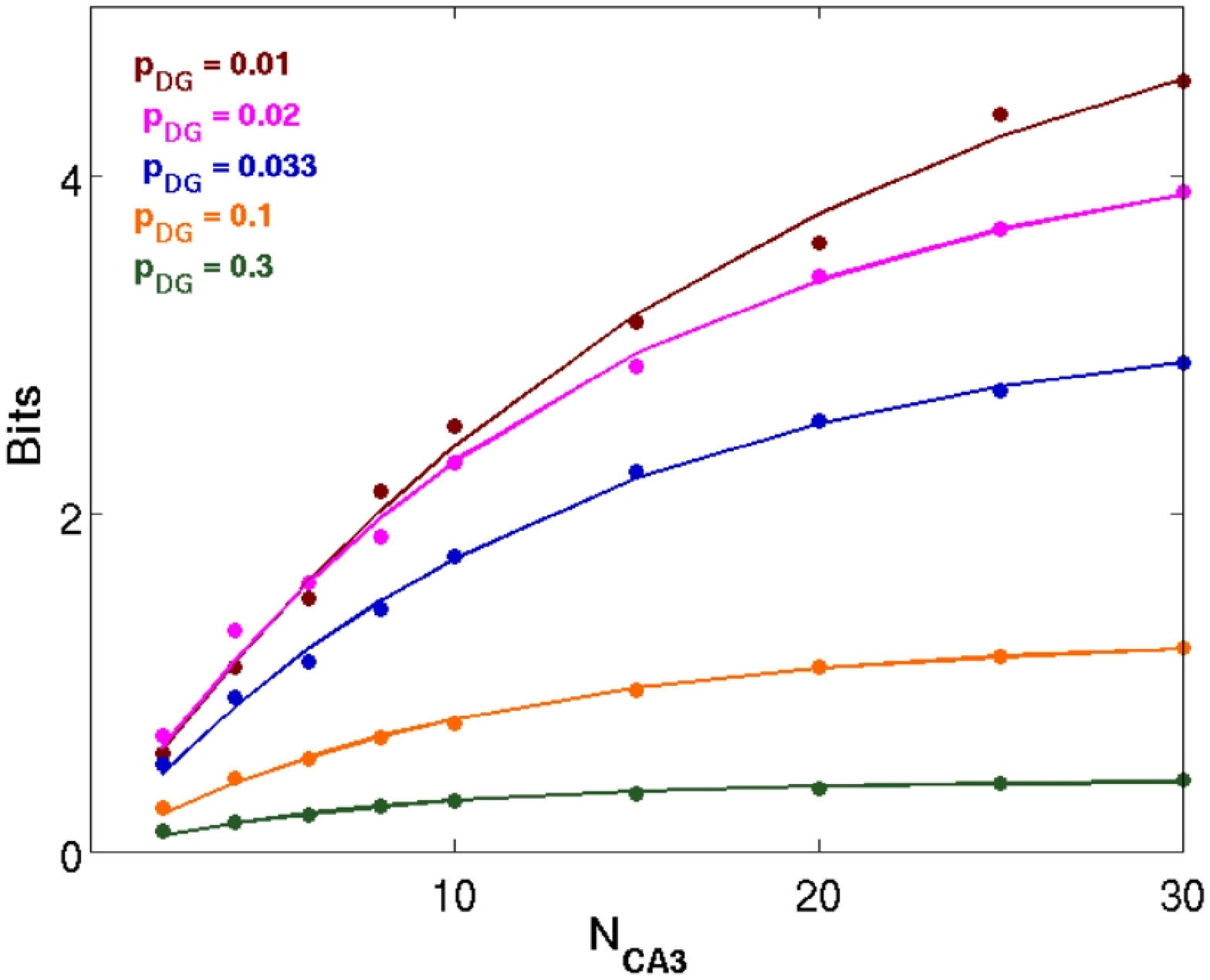}
\end{minipage}&
(b)\begin{minipage}{3.2in}
\includegraphics[width=\linewidth]{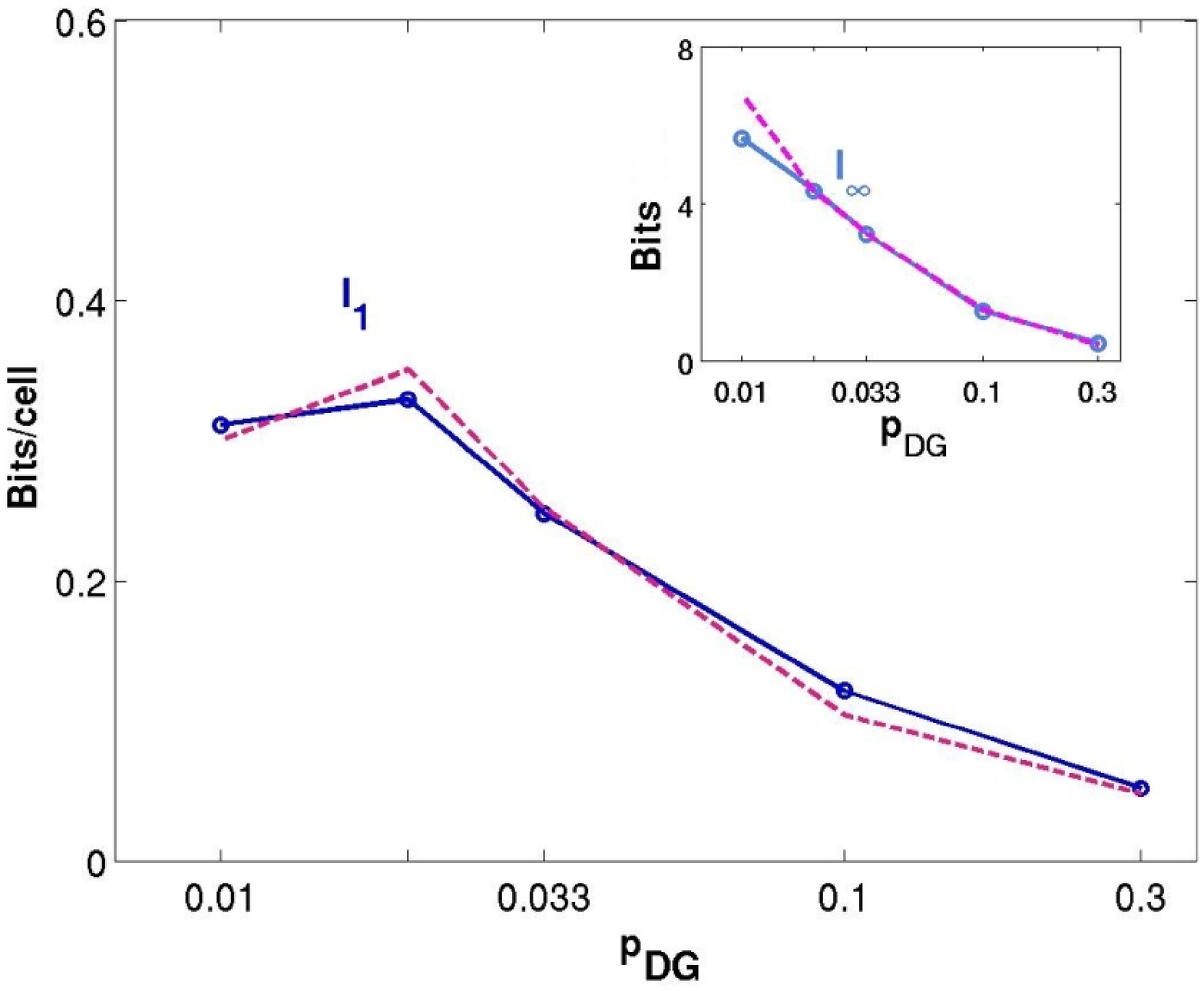}
\end{minipage}\\
\end{tabular}
\caption{{\bf Sparse DG activity is effective at driving CA3.} (a) Information plotted versus the number of CA3 units, different colors correspond to different values for $p_{DG}$. Dots represent information values obtained from simulations, while the curves are exponentially saturating fits to the data points, as described in Methods. (b) Plot of the two parameters of the fits. Main figure: slope parameter describing the slope of the linear part of the information curve (for low $N_{CA3}$); inset: total information parameter describing the saturation level reached by the information - both are contrasted with the corresponding measures (dashed lines) obtained varying $C_{MF}$ instead of $p_{DG}$.}\label{fig:6}
\end{figure*}

Results are simular to those obtained varying the sparsity of the MF connections. Indeed, the analytical estimate in the two conditions would be exactly the same, within the approximation with which we compute it, because the two parameters $p_{DG}$ and $C_{MF}$ enter the
calculation in equivalent form, as a product. The actual difference between the two parameters stems from the fact that increasing $C_{MF}$, CA3 units end up sampling more and more the same limited population of active DG units, while increasing $p_{DG}$ this population increases in size. This difference can only be appreciated from the simulations, which however show that the main effect remains the same: an information maximum for rather sparse DG activity (and sparse MF connections), The subtle difference between varying the two parameters can be seen better in the saturation information value: with reference to the standard case, in the center of the graph in the inset, to the right increasing $p_{DG}$ leads to more information than increasing $C_{MF}$, while to the left the opposite is the case, as expected.

\subsection{Full and simplified decoding procedures}
As noted above, we find that the analytical estimate of the information per unit is always considerably lower than the slope parameter of the fit to the measures extracted from the simulations, contrary to expectations since the latter require an additional decoding step, which implies some loss of information. We also find, however, that the measures of mutual information that we extract from the simulations are strongly dependent on the method used, in the decoding step, to construct the ``localization matrix", i.e. the matrix  which compiles the frequency with which the virtual rat was decoded as being in position $\vec{x}'$ when it was actually in position $\vec{x}$. All measures reported so far, from simulations, are obtained constructing what we call the {\em full} localization matrix $Q(\vec{x},\vec{x}')$ which, if the square environment is discretized into $20\times 20$ spatial bins, is a large $400\times 400$ matrix, which requires of order 160,000 decoding events to be effectively sampled. We run simulations with trajectories of 400,000 steps, and additionally corrected the information measures to avoid the limited sampling bias \cite{Treves95}.

An alternative, that allows extracting unbiased measures from much shorter simulations, is to construct a simplified matrix $\hat{Q}(\vec{x}-\vec{x}')$, which averages over decoding events with the same vector displacement between actual and decoded positions. $\hat{Q}(\vec{x}-\vec{x}')$ is easily constructed on the torus we used in all simulations, and being a much smaller $20\times 20$ matrix it is effectively sampled in just a few thousand steps. 

\begin{figure*}
\centering
\includegraphics[width=0.8\linewidth]{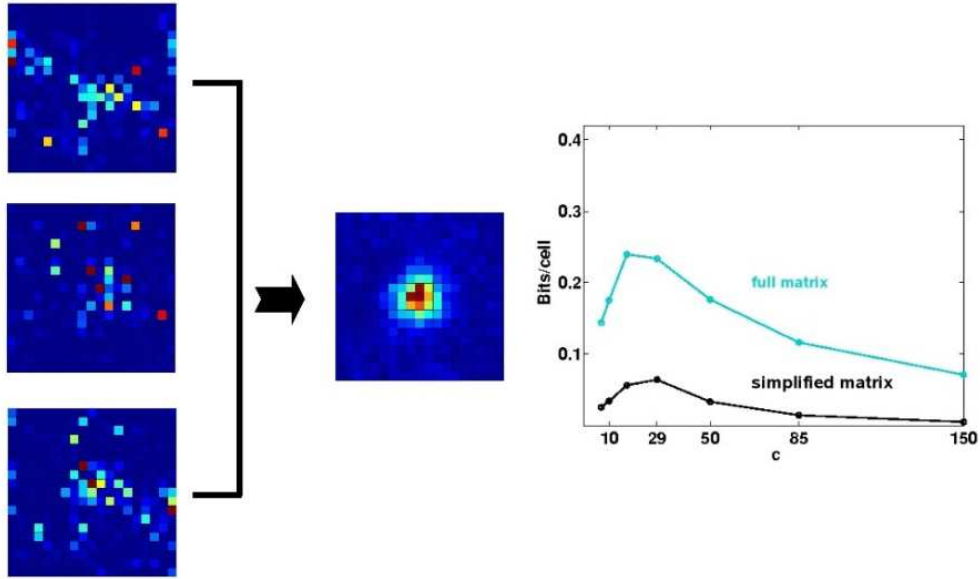}
\caption{{\bf Localization Matrices.} Left: the rows of the full matrix represent the actual positions of the virtual rat  while its columns represent decoded positions (the full matrix is actually $400\times 400$); three examples of rows are shown, rendered here as $20\times 20$ squares, all from decoding by a given sample of 10 units. The simplified matrix is a single $20\times 20$ matrix obtained (from the same sample) as the average of the full matrix taking into account traslation invariance. Right: the two procedures lead to large quantitative differences in information (here, the measures from samples of 10 units, divided by 10), but with the same dependence on $C_{MF}$.}\label{fig:7}
\end{figure*}

The two decoding procedures, given that the simplified matrix is the shifted average of the rows of the full matrix, might be expected to yield similar measures, but they do not, as shown in Fig.\ref{fig:7}. The simplified matrix, by assuming translation invariance of the errors in decoding, is unable to quantify the information implicitly present in the full distribution of errors around each actual position. Such errors are of an ``episodic" nature: the local view from position $\vec{x}$ might happen to be similar to that from position $\vec{x}'$, hence neural activity reflecting in part local views might lead to confuse the two positions, but this does not imply that another position $\vec{z}$ has anything in common with $\vec{z}+(\vec{x}'-\vec{x})$. Our little network model captures this discrepancy, in showing, in Fig.\ref{fig:7}, that for any actual position there are a few selected position that are likely to be erroneously decoded from the activity of a given sample of units; when constructing instead the translationally invariant simplified matrix, all average errors are distributed smoothly around the correct position (zero error), in a roughly Gaussian bell. 

Apparently, also the analytical estimate is unable to capture the spatial information implicit in such ``episodic" errors, as its values are well below those obtained with the full matrix, and somewhat above those obtained with the simplified matrix (consistent with some loss with decoding). These differences do not alter the other results of our study, since they affect the height of the curves,
not their shape, however they have important implications. The simplified matrix has the advantage of requiring much less data, i.e. less simulation time, but also less real data if applied to neurophysiological recordings, than the full matrix, and in most situations it might be the only feasible measure of spatial information (the analytical estimate is not available of course for real data). So in most cases it is only practical to measure spatial information with methods that, our model suggests, miss out much of the information present in neuronal activity, what we may refer to as ``dark information", not easily revealed. One might conjecture that the prevalence of dark information is linked to the random nature of the spatial code established by DG inputs. It might be that additional stages of hippocampal processing, either with the refinement of recurrent CA3 connections or in CA1, are instrumental in making dark information more transparent.

\subsection{Effect of learning on the Mossy Fibers}
\begin{figure}
\centering
\includegraphics[width=\linewidth]{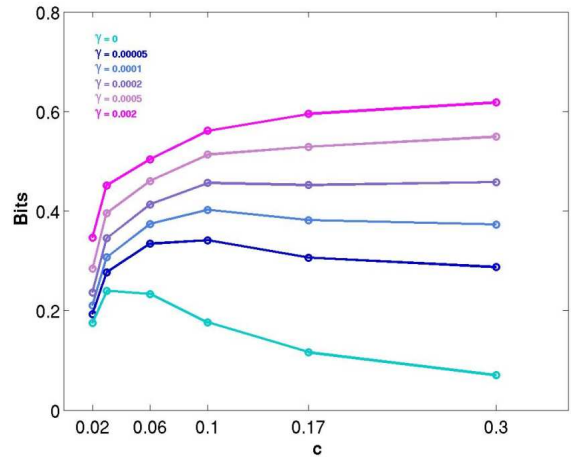}
\caption{{\bf Information vs. connectivity for different levels of learning.} Information is plotted as a function of the connectivity level between DG and CA3, different colors correspond to different values of the learning factor $\gamma_{MF}$. Simulations run for 100,000 training steps, during a fraction $\sim a_{CA3}=0.1$ of which each postsynaptic units is strongly activated, and its incoming weights liable to be modified. The $\gamma$ values tested hence span the range from minor modification of the existing weight, for $\gamma =0.00005$, to major restructuring of all available weights for $\gamma =0.002$.}\label{fig:8}
\end{figure}

While the results reported this far assume that MF weights are fixed, $J=1$, we have also conducted a preliminary analysis of how the amount of spatial information in CA3 might change as a consequence of plasticity on the mossy fibers. In an extension of the standard model, we allow the weights of the connections between DG and CA3 to change with a model ``Hebbian" rule. This is not an attempt to capture the nature of MF plasticity, which is not NMDA-dependent and might not be associative \cite{Nicoll05}, but only the adoption of a simple plasticity model that we use in other simulations. At each time step (that corresponds to a different place in space) weights are taken to change as follows:
\begin{equation}
\Delta J^{MF}_{ij}(t)= \gamma_{MF}\eta_{i}(\vec{x}(t))(\beta_{j}(\vec{x}(t))-<\beta(\vec{x}(t))>)
\end{equation}
where $\gamma_{MF}$ is a plasticity factor that regulates the amount of learning. Modifying in this way the MF weights has the general effect of increasing information values, so that they approach saturation levels for lower number of CA3 cells; in particular this is true for the information extracted from both full and simplified matrices.
In Fig.~\ref{fig:8}, the effect of such ``learning" is shown for different values of the parameter $\gamma_{MF}$, as a function of connectivity.
We see that allowing for this type of plasticity on mossy fibers leads to shift the maximum of information as a function of the connectivity level. The structuring of the weights effectively results in the selection of favorite input connections, for each CA3 unit, among a pool of availables ones; the remaining strong connections are a subset of those ``anatomically" present originally. It is logical, then, that starting with a larger pool of connnections, among which to pick the ``right" ones, leads to more information than starting with few connections, which further decrease in effective number with plasticity. We expect better models of the details of MF plasticity to preserve this main effect.

\begin{figure}
\centering
\includegraphics[width=0.8\linewidth]{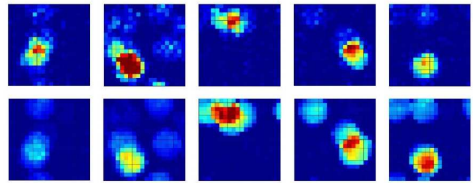}
\caption{{\bf MF plasticity can suppress, enlarge and in general refine CA3 place fields.} The place fields of five example units are shown before plasticity is turned on (top row) and after 100,000 steps with a large plasticity factor $\gamma_{MF}=0.0001$ (bottom row). The rounding and regularization of the fields was observed also for several other units in the simulation.}
\label{fig:9}
\end{figure}

A further effect of learning, along with the disappearance of some CA3 fields and the strengthening of others, is the refinement of their shape, as illustrated in Fig.\ref{fig:9}. It is likely that also this effect will be observed even when using more biologically accurate models of MF plasticity.

\subsection{Retrieval abilities}
\begin{figure}
\centering
\includegraphics[width=\linewidth]{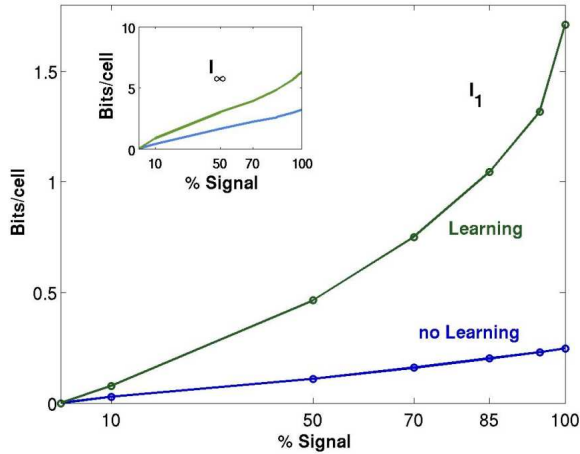}
\caption{{\bf Information reconstructed from a degraded input signal.} Slope parameter $I_{1}$ of the information curve as a function of the percentage of the DG input that CA3 receives. Inset: the same plot for the total information parameter $I_{\infty}$. The same training protocol was run as for Figs.~\ref{fig:8}-\ref{fig:9}.}
\label{fig:10}
\end{figure}

Finally, all simulations reported so far involved a full complement of DG inputs at each time step in the simulation. We have also tested the ability of the MF network to {\em retrieve} a spatial representation when fed with a degraded input signal, with and without MF plasticity. The input is degraded, in our simulation, simply by turning on only a given fraction, randomly selected, of the DG units that would normally be active in the environment. The information extracted after decoding by a sample of units (in Fig.~\ref{fig:10}, 10 units) is then contrasted with the size of the cue itself. In the absence of MF plasticity, there is obviously no real retrieval process to talk about, and the DG-CA3 network simply relays partial information. When Hebbian plasticity is turned on, the expectation from similar network models (see e.g. \cite{Treves04}, Fig.9) is that there would be some pattern completion, i.e. some tendency for the network to express nearly complete output information when the input is partial, resulting in a more sigmoidal input-output curve (the exact shape of the curve depends of course also on the particular measure used). 
It is apparent from Fig.~\ref{fig:10} that while, in the absence of plasticity, both parameters characterizing the information that can be extracted from CA3 grow roughly linearly with the size of the cue, with plasticity the growth is supralinear. This amounts to the statement that the beneficial effects of plasticity require a full cue to be felt -- the conceptual opposite to pattern completion, the process of integrating a partial cue using information stored on modified synaptic weights. This result suggests that the sparse MF connectivity is sub-optimal for the associative storage that leads to pattern completion, a role that current perspectives ascribe instead to perforant path and recurrent connections to CA3. The role of the mossy fibers, even if plastic, may be limited to the establishment of new spatial representations.

\section{Discussion}
Ours is a minimal model, which by design overlooks several of the elements likely to play an important role in the functions of the dentate gyrus - perhaps foremost, neurogenesis \cite{Kuhn}. Nevertheless, by virtue of its simplicity, the model helps clarify a number of quantitative issues that are important in refining a theoretical perspective of how the dentate gyrus may work.

First, the model indicates that the recently discovered multiplicity of place fields by active dentate granule cells \cite{Leutgeb07} might be just a ``fact of life", with no major computational implications for dentate information processing. Still, requiring that active granule cells express multiple fields seems to lead, in another simple network model (of how dentate activity may result from entorhinal cortex input \cite{Si09}), to the necessity of inputs from {\em lateral} EC, therefore to the refinement of sequential computational constraints on the operation of hippocampal circuits. 

Second, the model shows that, assuming a fixed total MF input strength on CA3 units, it is beneficial in information terms for the MF connectivity to be very sparse; but not vanishingly sparse. The optimal number of anatomical MF connections on CA3 units depends somewhat on the various parameters (the noise in the system, how sparse is the activity in DG and CA3, etc.) and it may increase
slightly when taking MF plasticity into account, but it appears within the range of the number, 46, reported for the rat by \cite{Amaral}. It will be interesting to see whether future measures of MF connectivity in other species correspond to those ``predicted" by our model once the appropriate values of the other parameters are also experimentally measured and inserted into the model. A similar set of consideration applies to the fraction of granule cells active in a given environment, $p_{DG}$, which in the model plays a similar, though not completely identical, role to $C_{MF}$ in determining information content.

Third, the model confirms that the sparse MF connections, even when endowed with associative plasticity, are not appropriate as devices  to store associations between input and output patterns of activity -- they are just too sparse. This reinforces the earlier theoretical view \cite{McMorris}, \cite{Treves92}, which was not based however on an analysis of spatial representations, that the role of the dentate gyrus is in establishing new CA3 representations and not in associating them to representations expressed elsewhere in the system. Availing itself of more precise experimental paramaters, and based on the spatial analysis, the current model can refine the earlier theoretical view and correct, for example, the notion that ``detonator" synapses, firing CA3 cells on a one-to-one basis, would be optimal for the mossy fiber system. The optimal situation turns out to be the one in which CA3 units are fired by the combination of a couple of DG input units, although this is only a statistical statement. Whatever the exact distribution of the number of coincident inputs to CA3, DG can be seen as a sort of {\em random pattern generator}, that sets up a CA3 pattern of activity without any structure that can be related to its anatomical lay-out \cite{Redish}, or to the identity of the entorhinal cortex units that have activated the dentate gyrus. As with random number generators in digital computers, once the product has been spit out, the exact process that led to it can be forgotten. This is consistent with experimental evidence that inactivating MF transmission or lesioning the DG does not lead to hippocampal memory impairments once the information has already been stored, but lead to impairments in the storage of new information \cite{Lassalle}, \cite{LeeKes}. The inability of MF connection to subserve pattern completion is also consistent with suggestive evidence from imaging studies with human subjects \cite{Bakker08}.

Fourth, and more novel, our findings imply that a substantial fraction of the information content of a spatial CA3 representation, over half when sampling limited subsets of CA3 units, can neither be extracted through the simplified method which assumes translation invariance, nor assessed through the analytical method (which anyway requires an underlying model of neuronal firing, and is hence only indirectly applicable to real neuronal data). This large fraction of the information content is only extracted through the time-consuming construction of the full localization matrix. To avoid the limited sampling bias \cite{Panzeri96} this would require, in our hands, the equivalent of a ten hour session of recording from a running rat (!), with a square box sampled in $20\times20$ spatial bins. We have hence labeled this large fraction as {\em dark information}, which requires a special effort to reveal. Although we know little of how the real system decodes its own activity, e.g. in downstream neuronal populations, we may hypothesize that the difficulty at extracting dark information affects the real system as well, and that successive stages of hippocampal processing have evolved to address this issue. If so, qualitatively this could be characterized as the representation established in CA3 being {\em episodic}, i.e. based on an effectively random process that is functionally forgotten once completed, and later processing, e.g. in CA1, may be thought to gradually endow the representations with their appropriate continous spatial character. Another network model, intended to elucidate how CA1 could operate in this respect, is the object of our on-going analysis.

The model analysed here does not include neurogenesis, a most striking dentate phenomenon, and thus it cannot comment on several intriguing models that have been put forward about the role of neurogenesis in the adult mammalian hippocampus \cite{Aimone06}, \cite{Becker05}, \cite{Wiskott06}. Nevertheless, presenting a simple and readily expandable model of dentate operation can facilitate the development of further models that address neurogenesis, and help interpret puzzling experimental observations. For example, the idea that once matured newborn cells may temporally ``label" memories of episodes occurring over a few weeks \cite{Kee07}, \cite{Ge07}, \cite{Buzzetti07}, \cite{Tashiro07} has been weakened by the observation that apparently even young adult-born cells, which are not that many \cite{Cameron01}, \cite{McDonald05}, \cite{Tashiro07}, are very sparsely active, perhaps only a factor of two or so more active than older granule cells \cite{Chawla}. Maybe such skepticism should be reconsidered, and the issue reanalysed using a quantitative model like ours. One could then investigate the notion that the new cells link together, rather than separating, patterns of activity with common elements (such as the temporal label). To do that clearly requires extending the model to include a description not only of neurogenesis, but also of plasticity within DG itself \cite{McHugh07} and of its role in the establishment of successive representations one after the other.

\begin{widetext}
\section{Methods}

\subsection{Replica calculation}
\label{replica}

\subsubsection*{Estimation of the equivocation}
Calculating the equivocation from its definition in Eq.\ref{equiv} is straightforward, thanks to the simplifying assumption of independent noise in CA3 units. We get
\begin{equation}
\langle H_{2}\left(\{\eta_{i}\}|\vec{x}\right)\rangle_{\vec{x}}=\frac{N}{\ln2}\int \frac{d\vec{x}}{A} \left\{ -\Phi(-\varrho_i(\vec{x}))\ln\Phi(-\varrho_i(\vec{x}))
+\Phi(\varrho_i(\vec{x}))\left[\frac{1}{2}+\ln(\sqrt{2\pi}\delta)\right] -\frac{1}{2}\varrho_i(\vec{x})\sigma(\varrho_i(\vec{x}))\right\}
\end{equation}
where 
\begin{eqnarray*}
\varrho_{i}&=&\frac{\bar{\eta_{i}}}{\delta}\\
\sigma (\varrho)&=& \frac{1}{\sqrt{2\pi}}e^{-\frac{\varrho^{2}}{2}},
\end{eqnarray*}
although the spatial integral remains to be carried out.

\subsubsection*{Estimation of the entropy}
For the entropy, Eq.\ref{entr}, the calculation is more complicated. Starting from
$$
H_{1}\left(\{\eta_{i}\}\right)=-\frac{1}{\ln2}\int\frac{d\vec{x}}{A}\prod_{i}d\eta_{i}\;P(\{\eta_{i}\}|\vec{x})\;\ln\left[\int \frac{d\vec{x}\prime}{A} P(\{\eta_{i}\}|\vec{x}\prime)\right]
$$
we remove the logarithm using the replica trick (see \cite{Mezard86})
\begin{equation}
H_{1}\left(\{\eta_{i}\}\right)=-\frac{1}{\ln2}\int\frac{d\vec{x}}{A}\prod_{i}d\eta_{i}\;P(\{\eta_{i}\}|\vec{x})\;\lim_{n \rightarrow 0}\frac{1}{n}\left\{\left[\int \frac{d\vec{x}\prime}{A} P(\{\eta_{i}\}|\vec{x}\prime)\right]^{n}-1\right\}
\end{equation}
which can be rewritten
\begin{equation}
H_{1}\left(\{\eta_{i}\}\right)=-\frac{1}{\ln2}\;\lim_{n \rightarrow 1}\frac{1}{(n-1)}\left[ \tilde{H_1}(n)-\tilde{H_1}(1)\right]
\label{hn}
\end{equation}
using the spatial averages, defined for an arbitrary real-valued number $n$ of replicas
\begin{eqnarray}
\tilde{H_1}(n)&\equiv&\int\prod_{i} d\eta_{i}\left[\int \frac{d\vec{x}}{A}P(\{\eta_{i}\}|\vec{x})\right]^{n}\nonumber\\
&=&\int \frac{d\vec{x}_{1}}{A}\dots \frac{d\vec{x}_{\beta}}{A}\dots \frac{d\vec{x}_{n}}{A}\prod^{N}_{i}h_{i}\left(\{\vec{x}_{\beta}\},n\right)
\end{eqnarray}
where we have defined a quantity dependent on both the number $n$ of replicas and on the position in space, later to be integrated over, of each replica $\beta$:
$$
h_{i}\left(\{\vec{x}_{\beta}\},n\right)\equiv\int{ d\eta_{i}\prod_{\alpha} \left[
\;\delta(\eta_{i})\;\Phi\left(- \frac{\bar{\eta}_{i}(\vec{x}_{\beta})} {\delta}\right) + \frac{1}{(\sqrt{2\pi}\delta)}\;
e^{-\frac{\left(\eta_{i}-\bar{\eta}_{i}(\vec{x}_{\beta}) \right)^{2}}{2\delta^{2}}}\;\Theta(\eta_{i})\right]}.
$$
We need therefore to carry out integrals over the firing rate of each CA3 unit, $\eta_i$, in order to estimate $h_{i}\left(\{\vec{x}_{\beta}\},n\right)$, while keeping in mind that in the end we want to take $n\to 1$. Carrying out the integrals yields a below-threshold and an above-threshold term
\begin{equation}
h_{i}\left(\{\vec{x}_{\beta}\},n\right)=\prod_{\beta}\Phi\left(-\varrho_{i}(\vec{x}_{\beta})\right) + e^{-(n-1)\frac{S}{2\delta^{2}}}\int_{-\infty}^{\eta_{0}}\frac{e^{-n\frac{\hat{\eta}^2}{2\delta^{2}}}}{(\sqrt{2\pi} \delta)^n} d\hat{\eta}
\end{equation}
where we have defined the quantities
\begin{eqnarray}
S(\{\vec{x}_{\beta}\})  &=& -\left[\frac{1}{n(n-1)}\left(\sum_{\beta}\bar{\eta_{i}}(\vec{x}_{\beta})\right)^{2}
-\frac{1}{(n-1)}\sum_{\beta}\bar{\eta_{i}}^{2}(\vec{x}_{\beta})\right]\nonumber\\
&=&\frac{1}{2n(n-1)}\sum_{\alpha,\beta}\left[\bar{\eta_{i}}(\vec{x}_{\alpha})-\bar{\eta_{i}}(\vec{x}_{\beta})\right]^{2}
\end{eqnarray}
and $\eta_{o} =(1/n)\sum_{\beta}\bar{\eta_{i}}(\vec{x}_{\beta})$, while $\hat{\eta}=\eta-\eta_{o}$.

One might think that $h_{i}\left(\{\vec{x}_{\beta}\},n\sim 1\right)\to 1 + (n-1)h_{i,n} \left(\{\vec{x}_{\beta}\} , 1\right) + O((n-1)^2 $, hence in the product over cells, that defines the entropy $H_{1}\left(\{\eta_{i}\}\right)$,
the only terms that survive in the limit $n\to 1$ would just be the summed single-unit contributions obtained from the
first derivatives with respect to $n$. This is not true, however, as taking the replica limit produces the counterintuitive effect that replica-tensor products of terms, which individually disappear for $n\to 1$, only vanish to first order in $n-1$, as shown by \cite{DelPrete01}. The replica method is therefore able, in principle, to quantify the effect of correlations among units, expressed in entropy terms stemming from the product of $h_i$ across units. 

Briefly, one has
\begin{eqnarray}
h_{i}\left(\{\vec{x}_{\beta}\},n\right) &=& \prod_{\beta}\Phi\left(-\varrho_{i}(\vec{x}_{\beta})\right) +
e^{-(n-1)\frac{S}{2\delta^{2}}}\left[\frac{1}{\sqrt{n}(\sqrt{2\pi}\delta)^{n-1}}\Phi\left(\sqrt{n}\eta_{o}/\delta\right)\right]\nonumber\\
&\simeq&\Phi\left(-\varrho_{o}\right)+\frac{1}{2}\frac{\sigma^{2}\left(-\varrho_{o}\right)}{\Phi\left(-\varrho_{o}\right)}\sum_{\alpha,\beta\neq\alpha}\left(\varrho(\vec{x}_{\alpha})-\varrho_{o}\right)\left(\varrho(\vec{x}_{\beta})-\varrho_{o}\right) \nonumber\\
&&+(n-1)\Phi\left(-\varrho_{o}\right)\ln\Phi\left(-\varrho_{o}\right)\nonumber\\
&&+\Phi\left(\varrho_{o}\right)+\frac{(n-1)}{2}\sigma\left(\varrho_{o}\right)\varrho_{o}
\nonumber\\
&&-(n-1)\Phi\left(\varrho_{o}\right)\left[\frac{1}{2}+\ln(\sqrt{2\pi}\delta)+\sum_{\alpha,\beta\neq\alpha}\frac{S(\vec{x}_{\alpha},\vec{x}_{\beta})}{2\delta^{2}}\right]
+(n-1)\sigma(\varrho_o)\frac{\partial\varrho_o}{\partial n}
\label{hpiccolo}\end{eqnarray}
where the first two rows come from the term below threshold, and the last two from the one above threshold. Then, following \cite{DelPrete01}, 
\begin{eqnarray}
H_{1}\left(\{\eta_{i}\}\right)\simeq-\frac{1}{\ln2}\;\lim_{n \rightarrow 1}\frac{1}{(n-1)}\left[\int \frac{d\vec{x}_{\alpha}}{A}\frac{d\vec{x}_{\beta}}{A}\left(1-(n-1)C+(n-1)\Gamma-\sum_{\alpha,\beta\neq\alpha}G_{\alpha,\beta}\right)^{N}-1\right]
\end{eqnarray}
where
\begin{eqnarray}
C&=&\Phi\left(\varrho_{o}\right)\left[\frac{1}{2}+\ln\left(\sqrt{2\pi}\delta\right)\right]-\frac{1}{2}\sigma\left(\varrho_{o}\right)\varrho_{o}
-\Phi\left(\varrho_{o}\right)\ln\Phi\left(-\varrho_{o}\right)\nonumber\\
\Gamma&=&\sigma\left(\varrho_{o}\right)\left[-\varrho(\vec{x}_1)+\frac{1}{n-1}\sum_{\beta >1}\varrho(\vec{x}_{\beta})\right]\label{CgammaG}\\
G_{\alpha,\beta}&=&\frac{\Phi\left(\varrho_{o}\right)}{4n(n-1)}\left(\varrho(\vec{x}_{\alpha})-\varrho(\vec{x}_{\beta})\right)^{2}+\frac{\sigma^{2}\left(-\varrho_{o}\right)}{2\Phi\left(-\varrho_{o}\right)}\left(\varrho(\vec{x}_{\alpha})-\varrho_{o}\right)\left(\varrho(\vec{x}_{\beta})-\varrho_{o}\right)\nonumber
\end{eqnarray}
and where we have considered that in the limit $n\to 1$ we have $\eta_{o}/\delta\equiv\varrho_o$ appear in all terms of finite weight. 

%RIVEDERE TUTTA QUESTA FORMULA E LE SEQUENTI, E RICONNETERSI ALLA EQUAZIONE (8).

The products between the matrices $G_{\alpha,\beta}$ attached to each CA3 unit generate the higher order terms in $N$. Calculating them in our case, in which different CA3 units can receive partially overlapping inputs from DG units, is extremely complex (see \cite{DelPrete02}, where information transmission across a network is also considered), and we do not pursue here the analysis of such higher order terms. One can retrieve the result of the TG model in Ref.~\cite{DelPrete01} by taking the further limit $\varrho_o\to 0$, which implies $\Phi(\varrho_o)\to 1/2$ and $\sigma^2(\varrho_o)\to 1/(2\pi)$. 
A further subtlety is that, in taking the $n\to 1$ limit, there is a single replica, say $\vec{x}$, which is counted {\em once} in the limit, but also several {\em different} replicas, denoted $\vec{x}\prime, \vec{x}\prime\prime, \dots $, whose weights vanish, but which remain to determine e.g. the terms proportional to $(n-1)$ emerging from the derivatives. Thus, in the very last term of Eq.~\ref{hpiccolo}, one has to derive $\varrho_o$ with respect to $n$ to produce the $\Gamma $ term of Eq.~\ref{CgammaG}, which is absent in \cite{DelPrete01} because it vanishes with $\varrho_o$. 
In the off-diagonal terms of the $G$ matrix there are $2(n-1)$ entries dependent on replicas $\vec{x}$ and $\vec{x}\prime$, and $(n-1)(n-2)$ entries dependent on replicas $\vec{x}\prime$ and $\vec{x}\prime\prime$. 

Focusing now solely on terms of order $N$, note that the term $S$ is effectively a {\em spatial signal}. In the $n\to 1$ limit it can be rewritten, using $\vec{x}$ for the single surviving replica, as 
$$ S(\vec{x},\vec{x}{\prime})= \left[\bar{\eta}_{i}\left(\vec{x}\right)
-\bar{\eta}_{i}\left(\vec{x}{\prime}\right)\right]^2-\frac{1}{2}\left[\bar{\eta}_{i}\left(\vec{x}\prime\right)
-\bar{\eta}_{i}\left(\vec{x}{\prime\prime}\right)\right]^2.
$$
This allows us to derive, to order $N$, our result for the spatial information content, Eq.~\ref{info-space}.

%\begin{eqnarray}
%H_{1}\left(\{\eta_{i}\}\right)\simeq\frac{N}{\ln2} & \left\{
%-\int \frac{d\vec{x}}{A} \Phi(-\varrho_i(\vec{x}))\ln \Phi(-\varrho_i(\vec{x}))+\int \frac{d\vec{x}}{A}\Phi(\varrho_i(\vec{x}))\left[\frac{1}{2}+\ln(\sqrt{2\pi}\delta)-\frac{1}{2}\varrho_i(\vec{x})\sigma(\varrho_i(\vec{x})\right] \right. \nonumber\\
%&+\left.\int \frac{d\vec{x}}{A}\frac{d\vec{x}{\prime}}{A}\frac{S(\vec{x},\vec{x}{\prime})}{2\delta^2}+\int \frac{d\vec{x}}{A}\frac{d\vec{x}{\prime}}{A}\left[\frac{\sigma^{2}\left(\varrho_{o}\right)}{\Phi\left(-\varrho_{o}\right)}\left(\varrho(\vec{x})-\varrho_{o}\right)\left(\varrho(\vec{x}{\prime})-\varrho_{o}\right)\right]\right\}
%\end{eqnarray}
 
Note that when the threshold of each unit tends to $-\infty$, and therefore its mean activation $\varrho_{oi}\to\infty $, our units behave as threshold-less linear units with gaussian noise, and the information they convey tends to 
\begin{equation}
\langle I\left(\vec{x},\{\eta_{i}\}\right)\rangle=\frac{N}{4\ln2}\langle \int \frac{d\vec{x}}{A}\frac{d\vec{x}{\prime}}{A}\left[\varrho_{i}\left(\vec{x}\right)
-\varrho_{i}\left(\vec{x}{\prime}\right)\right]^2\rangle
\end{equation}
which is simply expressed in terms of a spatial signal-to-noise ratio, and coincides with the results in Refs. \cite{Samengo}, \cite{DelPrete01}.

\subsection{$m$-field Decomposition}
\label{decomposition}

Eqs.~\ref{info-space} and \ref{info-space1cell} simply sum equivalent average contributions from each CA3 unit. Each such contribution can then be calculated as a series in $m$, the number of DG fields feeding into the CA3 unit. One can in fact write, for example,
\begin{eqnarray*}
\langle\Phi(\varrho_i(\vec{x}))\varrho_i^2(\vec{x})\rangle&=& 
\frac{P(1)}{\delta^2}\left\{\sum_{Q_{1}=0}^{\infty} P(Q_{1})\Phi(\varrho_i(\vec{x}))\left[ \sum_{j}J\beta_{j}\left(\vec{x},\{\vec{x}_{jk}\}\right)-T\right]^{2}\right\}
\\
&+&\frac{P(2)}{\delta^2}\left\{\sum_{Q_{1},Q_{2}=0}^{\infty} P(Q_{1})P(Q_{2})\Phi(\varrho_i(\vec{x})) \left[\sum_{j} J\beta_{j}\left(\vec{x},\{\vec{x}_{jk}\}\right)-T\right]^{2}\right\}
\\
&+&\dots + \frac{P(\gamma)}{\delta^2} \left\{\sum_{Q_{1},Q_{2}\dots Q_{\gamma}=0}^{\infty} P(Q_{1})P(Q_{2})\dots P(Q_{\gamma})\Phi(\varrho_i(\vec{x}))\right.\\
&&\phantom{+\dots + P(\gamma)}\left.
\left[\sum_{j}J \beta_{j}\left(\vec{x},\{\vec{x}_{jk}\}\right)-T\right]^{2}\right\}+\dots
\end{eqnarray*}
where in each term there are $\gamma$ active DG units, indexed by $j$, presynaptic to CA3 unit $i$, and each has $Q_j$ fields
(including the possibility that $Q_j=0$), indexed by $k$. 
\end{widetext}

A similar expansion can be written for the other terms.
One then realizes that the spatial component reduces to integrals that depend solely on the total number of fields $m=\sum_j^n Q_j$, no matter how many DG active units they come from, and the expansion can be rearranged into an expansion in $m$
\begin{equation}
\langle I\left(\vec{x},\{\eta_{i}\}\right)\rangle\, =\frac{N}{2\ln2}\sum_{m=0}^{\infty}C_mD_m(T)
\end{equation}
where one of the components in each term is, for example,
\begin{equation}
D_m(T)= \int \frac{d\vec{x}}{A}\frac{d\vec{x}\prime}{A} \frac{d\vec{x}_1}{A}\dots\frac{d\vec{x}_m}{A} \Phi(\varrho(\vec{x},\{\vec{x}_{j}\}))\varrho^{2}\left(\vec{x}{\prime},\{\vec{x}_{j}\}\right)
\end{equation}
with $\varrho(\vec{x}{\prime},\{\vec{x}_{j}\})=\left[J\sum_{l=1}^{m} \psi\left(\vec{x}{\prime}-\vec{x}_{j}\right)-T\right]/\delta $ the mean signal-to-noise at position $\vec{x}$ produced by $m$ fields, from no matter how many DG units. The numerical coefficient $C_m$, instead, stems from the combination of the distribution for the number of fields for each presynaptic DG unit active in the environment, which differs between models A, B and C, and the Poisson distribution for the number of such units 
\begin{eqnarray*}
P\left(\gamma\right)&=& \frac{\left(\alpha\right)^{\gamma}}{\gamma!}e^{-\alpha}\\
\alpha&=&p_{DG}C_{MF}.
\end{eqnarray*}
The sum extends in principle to $m\to\infty$, but in practice it can be truncated after checking that successive terms give vanishing contributions. The appropriate truncation point obviously depends on the mean number of fields $q$, as well as on the model distribution of fields per unit. Note that the first few terms (e.g. for $m=0,1,\dots $) may give negative but not necessarily negligible contributions if the effective threshold $T$ is high.

For model A, 
$$
P_A\left(Q\right)=\frac{q^{Q}}{Q!}e^{-q}
$$
and combining the two Poisson series one finds
\begin{equation}
C_m=e^{\alpha\left(e^{-q}-1\right)}K_m\frac{q^{m}}{m!}
\end{equation}
where $K_0\equiv 1$ and the other $K_m(\lambda ) $ are the polynomials
$$
\left\{
\begin{array}{rl}
K_1 &= \lambda \\
K_{2} &=\lambda+\lambda^{2}\\
K_{3} &=\lambda+3\lambda^{2}+\lambda^{3}\\
K_{4} &=\lambda+7\lambda^{2}+6\lambda^{3}+\lambda^{4}\\
\vdots & \\
K_m & = \sum_{l=1}^m \; \Gamma(l,m)  \; \lambda^{l} \\
\vdots
\end{array}
\right.
$$
given by the modified Khayyam-Tartaglia recursion relation $$\Gamma (l,m) = \Gamma (l-1,m-1)+l\;\Gamma (l,m-1)$$ and where $\lambda=\alpha e^{-q}$.

For model B, 
$$
P_B(Q)=\frac{1}{1+q}\left(\frac{q}{1+q}\right)^{Q_i}
$$
and combining the Poisson with the exponential series one finds
\begin{equation}
C_m=e^{\alpha\left(\frac{1}{1+q}-1\right)}\tilde{K}_m\left(\frac{q}{1+q}\right)^{m}
\end{equation}
where again $\tilde{K}_0\equiv 1$, while the other $\tilde{K}_m(\tilde{\lambda} ) $ are the distinct polynomials
$$
\left\{
\begin{array}{rl}
\tilde{K}_1 &= \tilde{\lambda} \\
\tilde{K}_{2} &=\tilde{\lambda}+\tilde{\lambda}^{2}/2!\\
\tilde{K}_{3} &=\tilde{\lambda}+2\tilde{\lambda}^{2}/2!+\tilde{\lambda}^{3}/3!\\
\tilde{K}_{4} &=\tilde{\lambda}+3\tilde{\lambda}^{2}/2!+3\tilde{\lambda}^{3}/3!+\tilde{\lambda}^{4}/4!\\
\vdots & \\
\tilde{K}_m & = \sum_{l=1}^m \; \tilde{\Gamma}(l,m)  \; \tilde{\lambda}^{l} \\
\vdots
\end{array}
\right.
$$
given by the further modified Khayyam-Tartaglia recursion relation 
$$\tilde{\Gamma} (l,m) = \tilde{\Gamma} (l-1,m-1)/l+\tilde{\Gamma} (l,m-1)$$ and where $\tilde{\lambda}=\alpha/(1+q)$.

For model C, 
$$
P_C(Q)=\delta_{1Q}
$$
there is no parameter $q$ (i.e., $q\equiv 1$), and one simply finds
\begin{equation}
C_m=e^{-\alpha}\frac{\alpha^m}{m!}.\label{C_C}
\end{equation}
Note that in the limit $q\to 0$, when the mean input per CA3 unit $\mu=\alpha q$ remains finite, for both models A and B one finds
$$
\lim_{q\to 0} C_m= e^{-\mu}\frac{\mu^m}{m!}
$$
which is equivalent to Eq.~\ref{C_C}, in line with the fact that both models A and B reduce, in the $q\to 0$ limit, to single-field distributions, but even units with single fields become vanishingly rare, so formally one has to scale up the mean number of active presynaptic units, $\alpha$, to keep $\mu\equiv\alpha q$ finite and establish the correct comparison to model C.

\subsection{Sparsity and Threshold}
\label{sparsity}
\begin{figure}
\centering
\includegraphics[width=\linewidth]{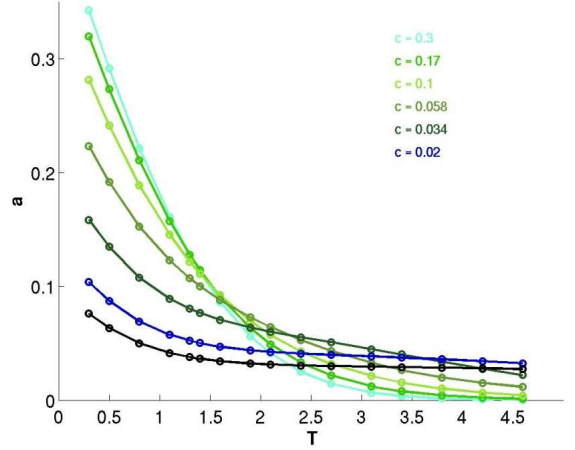}
\caption{{\bf Sparsity-Threshold relation.} The sparsity $a$ of CA3 layer vs. the threshold $T$ of CA3 units, from the numerical integration of Eq.~\ref{spars-thresh}. Different lines correspond to different degrees of connectivity between DG and CA3.}\label{fig:11}
\end{figure}

The analytical relation between the threshold $T$ of CA3 units and the sparsity $a$ of the layer is obtained starting from the formula defining formula the sparsity $a_{CA3}\equiv\frac{\langle\eta_{i}\left(\vec{x}\right)\rangle^{2}}{\langle\eta_{i}^{2}\left(\vec{x}\right)\rangle}$ which can be rewritten
\begin{equation}
a=\langle\frac{\left[\sigma\left(\varrho(\vec{x})\right)+\varrho(\vec{x})\Phi\left(\varrho(\vec{x})\right)\right]^{2}}{\left[\varrho(\vec{x})\sigma\left(\varrho(\vec{x})\right)+\Phi\left(\varrho(\vec{x})\right)\left(1+\varrho^{2}(\vec{x})\right)\right]}\rangle
\label{spars-thresh}
\end{equation}
Since in the analytical calculation we have $T$ as parameter, this equation can be taken as a relation $a(T)$ which has to be inverted to allow a comparison with the simulations, which are run controlling the sparsity level at a predefined level (in our case $a=0.1$) and adjusting the threshold parameter accordingly. The inversion requires using the $m$-field decomposition and numerical integration. A graphical example of the numerical relation is given in Fig.~\ref{fig:11}.

\subsection{Simulations}
The mathematical model described above was simulated with a network of 500 DG cells and 500 CA3 cells. A virtual rat explores a continuous two dimensional space, intended to represent a $1sqm$ square environment but realized as a torus, with periodic boundary conditions. For numerical purposes (estimating mutual information), the environment is discretized in a grid of $20\times 20$ locations, whereas trajectories are in continuos space, but in discretized time steps. In each time step (intended to correspond to roughly $62.5ms$, half a theta cycle, the virtual rat moves half a grid unit ($2.5cm$) in a direction similar to the direction of the previous time step, with a small amount of noise. To allow construction of a full localization matrix with good statistics, simulations are run for typically 400,000 time steps (while for the simplified translationally invariant matrix 5,000 steps would be sufficient). The space has periodic boundary conditions, as in a torus, to avoid border effects; the longest possible distance between any two locations is hence equal to 14.1 grid units, or $70cm$.

\subsubsection{DG place fields}
After assigning a number of firing fields for each DG units, according to the distributions of models A, B and C, we assign to each field a randomly chosen center. The shape of the field is then given by a Gaussian bell with that center. The tails of the Gausssian function are truncated: the value of the function is set to zero when the distance from the center is larger than a fixed radius $r=\sqrt{\frac{fA}{\pi}}$, with $f=0.1$ the ratio between the area of the field and the environment area $A$. In the standard model, only about 3 percent of the DG units are active in the environment, in agreement with experimental findings \cite{Chawla}; i.e. the DG firing probability is $p_{DG}=0.033$. The firing of DG units is not affected by noise, nor by any further threshold. Peak firing is conventionally set, in the center of the field, at the value $\frac{r^2}{2\pi}=2.02$, but DG units can fire at higher levels if they are assigned two or more overlapping fields.

\subsubsection{CA3 activation}
CA3 units fire according to Eq.~\ref{firetaequ}: the firing of a CA3 unit is a linear function of the total incoming DG input, distorted by a noise term. This term is taken from a gaussian distribution centered on zero, with variance $\delta=1$, and it changes for each unit and each time step. A threshold is imposed in the simulations to model the action of inhibition, hypothesizing that it serves to adjust the sparsity $a$ of CA3 activity to its required value. The sparsity is defined as
$$
a=\frac{\langle\eta_{i}\left(\vec{x}\right)\rangle^2}{\langle\eta_{i}^{2}\left(\vec{x}\right)\rangle}
$$
and it is fixed to $a=0.1$. This implies that the activity of the CA3 cells population is under tight inhibitory control.

\subsubsection{The decoding procedure and information extraction}
\label{decoding}
At each time step, the firing vector of a set of CA3 units is compared to all the average vectors recorded at each position in the $20\times 20$ grid, for the same sample, in a test trial (these are called template vectors). The comparison is made calculating the Euclidean distance between the current vector and each template, and the position of the closest template is taken to be the decoded position at that time step, for that sample. This procedure has been termed maximum likelihood Euclidean distance decoding \cite{EDdecoding}. The frequency of each pair of decoded and real positions are compiled in a so-called ``confusion matrix", or localization matrix, that reflects the ensemble of conditional probabilities $\left\{P\left(\left\{\eta_{i}\right\}|\vec{x}\right)\right\}$ for that set of units. Should decoding ``work" in a perfect manner, in the sense of always detecting the correct position in space of the virtual rat, the confusion matrix would be the identity matrix. 
From the confusion matrix obtained at the end of the simulation, the amount of information is extracted, and plotted versus the number of CA3 units present in the set. We averaged extensively over CA3 samples, as there are large fluctuations from sample to sample, i.e. for each given number of CA3 units we randomly picked several different groups of CA3 units and then averaged the mutual information values obtained. In all the results reported we averaged also over 3-4 simulation run with a different random number generator, i.e. over different trajectories. The same procedure leading to the information curve was repeated for different values of the parameters.
In all the information measures we reported, we also corrected for the limited sampling bias, as discussed by \cite{Treves95}.
In our case of spatial information, the bias is essentially determined by the spatial binning we used ($20\times 20$) and by the decoding method \cite{Panzeri99}.

\subsubsection{Fitting}
We fit the information curves obtained in simulations as a function of $N$ in order to get the values of the two most relevant parameter that describe their shape: the initial slope $I_{1}$ (i.e. the average information conveyed by the activity of individual units) and the total amount of information $I_{\infty}$ (i.e. the asymptotic saturation value). The function we used for the fit is the following
\begin{equation}
F(N)=I_{\infty}\left(1-e^{-N\frac{I_{1}}{I_{\infty}}}\right)
\end{equation}
In most cases the fit was in excellent agreement with individual data points, as expected on the basis of previous analyses \cite{Samengo}.

\subsection*{Acknowledgements}

We had valuable discussion with Jill Leutgeb, Bailu Si and Federico Stella.

%\section{Figure Legends}

%\subsection*{Figure 1}

\bibliographystyle{plain}      % number
\bibliography{DG}

\end{document}